%% file: paper.tex
\documentclass[conference]{IEEEtran}
\IEEEoverridecommandlockouts
\input{header}
\usepackage[all]{nowidow}
\usepackage{comment}
\input{authors_conf}

\usepackage{varwidth}
\newcommand\Umbruch[2][1.3cm]{\begin{varwidth}{#1}\centering#2\end{varwidth}}

\begin{document}

\maketitle

\begin{abstract}
In this paper, we propose a novel channel estimation scheme for pulse-shaped multicarrier systems using smoothness regularization for \gls{urllc}.
It can be applied to any multicarrier system with or without linear precoding to estimate challenging doubly-dispersive channels. 
A recently proposed modulation scheme using orthogonal precoding is \gls{otfs}.
In \gls{otfs}, pilot and data symbols are placed in \gls{dd} domain and are jointly precoded to the \gls{tf} domain.
On the one hand, such orthogonal precoding increases the achievable channel estimation accuracy
and enables high \gls{tf} diversity at the receiver.  
On the other hand, it introduces leakage effects which requires extensive leakage suppression when the piloting is jointly precoded with the data.
To avoid this, we propose to precode the data symbols only, place pilot symbols without precoding into the \gls{tf} domain, and estimate the channel coefficients by interpolating smooth functions from the pilot samples.
Furthermore, we present a piloting scheme enabling a smooth control of the number and position of the pilot symbols. 
Our numerical results suggest that the proposed scheme provides accurate channel estimation with reduced signaling overhead compared to standard estimators using Wiener filtering in the discrete \gls{dd} domain. 
\end{abstract}
\section{Introduction \label{sec:introduction}}
\glsresetall
Future mobile multicarrier systems have to meet a large variety of requirements. 
They are driven by increasingly demanding applications. 
Especially, the connectivity of high mobility devices such as automated vehicles poses a challenge.
Automated vehicles have very strict requirements regarding the quality of the communication which is commonly referred to as \gls{qos} \cite{kousaridas2021qos}. 
In particular, \gls{urllc} plays an important role in this context \cite{li20185g}.
\textcolor{black}{It is essential for automated vehicles that sufficient \gls{qos} parameters, such as latency and data rate, are reliably provided and this even in high mobility scenarios.}
In these scenarios, the wireless channel is considered to be doubly-dispersive, i.e., varying in both time and frequency.
In addition, efficiency plays an essential role due to limitations of the available spectrum as it is already foreseen that the \gls{5g} cannot fulfill future spectrum needs \cite{Euler2021}. 
For this reason, it is important to aim at improved efficiency during the development of future mobile multicarrier systems.
It does not suffice to focus exclusively on improvements at higher layers; the physical layer must also be addressed.
For example, it is desirable to reduce signaling overhead, e.g., the number of pilot signals, and to increase the reliability of the multicarrier system to avoid packet retransmissions.
In this paper, we focus on physical layer enhancements by proposing a novel channel estimation scheme and utilizing linear precoding.
\par
To address those challenges, we need to improve the transceiver structure of multicarrier systems taking pulse-shaping filters into account.  
Nowadays, \gls{ofdm} is broadly used, e.g., in the \gls{4g}, \gls{5g}, and \gls{wifi}.
\gls{ofdm} uses rectangular pulses at the transmitter and receiver filterbank.
With this setup, time-invariant channels reduce to convolution operators which are easily manageable, but \gls{ofdm} suffers significant performance losses, when the channel is time-variant \cite{jung:ieeecom:timevariant, wang2006performance}.  
In this context, \gls{otfs} has been introduced by Hadani et. al. \cite{hadani2017orthogonal}. 
It uses the \gls{dsft} as orthogonal precoding transform 
to precode symbols over the entire \gls{tf} domain.
This approach is very distinct as data and pilot symbols are both placed in the \gls{dd} domain and are jointly orthogonal precoded \cite{raviteja2019embedded}. 
Several studies show that \gls{otfs} significantly outperforms \gls{ofdm} in terms of \gls{ber} performance \cite{raviteja2018practical,Nimr2018,PfadlerWSA2020,gaudio2021otfs}. 
This is due to the fact that the joint orthogonal precoding enables high \gls{tf} diversity.
In particular the achievable channel estimation accuracy is increased, since a pilot symbol placed in the \gls{dd} domain probes each \gls{tf} coefficient \cite{raviteja2018interference,PfadlerICC}.  
However, it also comes with some disadvantages.
Firstly, channel estimation suffers under leakage effects when it is done in the discrete \gls{dd} domain \cite{PfadlerWCL}.
Secondly, resource allocation becomes less flexible regarding multiuser aspects \cite{Augustine2019}.
Thirdly, the overhead for piloting in the uplink grows proportionally to the number of users \cite{Thomas2022}.
This motivates the approach followed in this paper, which is to 
apply precoding to the data but not the pilot symbols. 
Although we loose some \gls{tf} diversity this way, we gain the flexibility to choose any precoding for the data symbols without affecting the piloting scheme. 
In \cite{zemen2018iterative}, it is shown that aside from the \gls{dsft} any other orthogonal precoding, i.e., 2D orthogonal transform, yields the same high \gls{tf} diversity, e.g., the low-complexity \gls{2D-fwht}.
\par
In particular, the estimation of doubly-dispersive channels is a very important aspect for future multicarrier systems especially when \gls{tf} symbols are precoded.
Since the provision of an accurate \gls{csi} and the usage of an appropriate equalizer is essential to enable high \gls{tf} diversity gains. 
Vehicular channels are considered to be doubly-dispersive, underspread, and often also to be sparse in the continuous \gls{dd} domain following the \gls{wssus} model \cite{bello1963characterization}.
A channel is underspread if all delay shifts and Doppler shifts are contained within a small region, i.e., both are relatively small. 
The channel is sparse when only a few point-like scatterers in the continuous \gls{dd} domain exist.
For pulse-shaped multicarrier filterbanks, the inherent sparsity of the channel cannot be harnessed using any form of \gls{dft}
for channel estimation in the discrete \gls{dd} domain \cite{Seo2010, Xiong2013, PfadlerWCL}.
A common way to estimate the channel is to get the \gls{ls} estimator from the pilot samples and to smooth them by means of the Wiener filtering in the discrete \gls{dd} domain, which is commonly referred to as \gls{lmmse} estimator or Markov estimator \cite{hoeher1997two,savaux2017lmmse}.
This approach however suffers under leakage effects \cite{taubock2010compressive,Xiong2013}. 
Leakage effects are caused by the presence of both fractional Doppler shifts and fractional delay shifts which are not consistent with the discrete nature of the \gls{dsft} \cite{PfadlerWCL}.
\par
To cope with leakage effects and to promote sparsity, more 
complex estimation schemes are commonly followed.
In this scope, compressed sensing or even super resolution are possible schemes, see for example \cite{taubock2010compressive,beinert2021super}, respectively.
In \cite{Rasheed2020}, Rasheed 
et al. propose a compressed sensing based algorithm using orthogonal
matching and modified subspace pursuit to estimate the time-varying channels.
A framework for sparse Bayesian learning with Laplace priors and a new piloting scheme has been introduced by Zhao et al. in \cite{Zhao2020}, where they consider fractional Doppler shifts but not fractional delay shifts. 
An off-grid sparse signal recovery to estimate the original channel rather than the effective discrete channel in the \gls{dd} domain is proposed in \cite{Wei2022}. 
In \cite{Liu2021}, an iterative optimization method is presented by Liu et. al., where a message passing signal recovery algorithm is utilized for channel estimation which takes
fractional Doppler shifts but not fractional delay shifts into account.
The listed schemes are rather complex, require high computing power, and consider longer time intervals, e.g., are computed adaptively over multiple frames,
which does not suite well to \gls{urllc} in the context of rapidly changing vehicular channels, as it is known that the \gls{wssus} assumption only holds for a limited duration and bandwidth \cite{Bernado2013}. 
This makes channel estimation challenging and requires channel estimation on a per frame basis \cite{PfadlerWSA2020}.
Computationally complex and iterative optimization methods are therefore not considered in the presented paper. 
\par
Focusing on low-complexity estimators for \gls{urllc}, a common choice is the estimation of the \gls{cmd} on a per frame basis. 
This can be done by using an \gls{lmmse} estimator which however suffers from leakage effects.
In this paper, we propose a novel \gls{cmd} estimator in the \gls{tf} domain in contrast to the estimation in the \gls{dd} domain used for \gls{otfs} and \gls{dft} based schemes for \gls{ofdm}. 
We place pilot symbols in the \gls{tf} domain to enable higher flexibility and reduced overhead for pilot signaling. 
However, the pilot and data symbols still need to be properly arranged within a rectangular frame. 
To apply fast orthogonal precoding transformations, we typically require the input dimension to be to the power of two which equals the number of data symbols. 
Therefore, the placement of the pilot symbols is not obvious. 
To control the number and position of the pilot symbols, we propose an algorithm and a so called \emph{accordion pilot placement} to place pilots in between the precoded symbols in the \gls{tf} domain.
The main contributions of this paper can be summarized as follows:
\begin{itemize}
\item We study pulse-shaped multicarrier systems with linear precoding for \gls{urllc} over doubly-dispersive channels,
\item we numerically compare different linear precoding transformations, 
\item we propose a novel smoothness optimized estimation scheme of the \gls{cmd} coefficients which minimizes the energy of the discrete Hessian and takes the ratio between the delay spread and Doppler spread, 
 the self-interference power, and receiver noise into account, and
\item we introduce a pilot placement scheme, i.e., \emph{accordion pilot placement}, 
which enables a smooth control of the number and position of the pilot symbols. 
\end{itemize}
\textcolor{black}{\subsection{Paper Organization}}
In Section~\ref{section-system-model}, the Gabor signaling and doubly-dispersive channel model is introduced. 
Linear precoding transforms and their diversity gain are discussed in Section~\ref{sec-spreading-diversity}.
\textcolor{black}{In Section ~\ref{section-channel-estimator}, we detail channel estimation, leakage effects, equalization, data recovery, and the proposed channel estimation scheme.}
The accordion pilot placement is presented in Section~\ref{section-accordion}.
In Section~\ref{section-numerics}, we show our numerical results. 
Finally, we summarize our conclusions in Section~\ref{sec:conclusions}.
\textcolor{black}{\subsection{Notational Remarks}
Random variable vectors, 2D-arrays and matrices are denoted with bold letters. 
Superscripts $(\cdot)^*$ and $(\cdot)^H$ denote the complex conjugate and the Hermitian transpose, respectively. 
Let $\ast$ denote the non-cyclic 2D convolution which only returns the valid part.
The column-wise vectorization operator, the absolute value, the euclidean norm, and the Frobenius norm is denoted as $\vectorize\{\cdot\}$, $|\cdot |$, $\|\cdot \|_2$, and $\|\cdot\|_\text{F}$, respectively.
We denote $\delta(\cdot)$  as the Dirac distribution, $\odot$ as the \textit{Hadamard product}, $\mathbb{E}\{\cdot\}$ as expectation operation, and $j^2=-1$.
We denote the indices of down-converted received signal by $(\bar{\cdot})$.} 
\section{System Model}
\label{section-system-model}
In this section, we introduce the system model which includes the doubly-dispersive channel and the input-output mapping of the information resources. 
We use a time-continuous Gabor (Weyl-Heisenberg) signaling to derive a discrete system model for the pulse-shaped multicarrier scheme.
We define the Gabor grid $\Lambda = F \Z_M \times T \Z_N$ with frequency step size $F>0$ and time step size $T>0$.
The indices $\mathcal I = \Z_M \times \Z_N$ run over the cyclic groups $\Z_M = \Z / M \Z$ (integers of modulo $M$) and $\Z_N = \Z / N\Z$ (integers of modulo $N$) taking in total $M$ frequency steps and $N$ time steps into account.
The overall frame duration $T_f$ and bandwidth $B$ are given by the products $TN$ and $FM$, respectively.
Regular Gabor grids can be categorized into three types depending on their 
\gls{tf} product $TF$: Oversampling if $TF < 1$, critical sampling for $TF=1$, and undersampling if $TF> 1$.
\par
Let us denote the complex-valued pulse-shaping filters for synthesis and analysis as $\gamma(t)$ and $g(t)$, respectively.
We design the pulses to be biorthogonal to \textcolor{black}{obtain a} perfect reconstruction in the absence of noise and channel distortions, i.e.,
\begin{equation}
   \int  g(t)^\ast \gamma(t-nT) e^{2 \pi j mF t} \d t =
     \begin{cases}
        1, & m = n = 0\\
        0, & \text{else}
    \end{cases},
    \label{eq:pulse}
\end{equation}
\textcolor{black}{At the receiver the orthogonality is typically lost due to channel dispersion which in turn causes self-interference \cite{kozek1996matched,kozek:nofdm,liu2004orthogonal,jung:ieeecom:timevariant,jung2007weyl,jung2007wssus,PfadlerWSA2020}.}
%
\par
At the transmitter, the Gabor filterbank uses the synthesis pulse $\gamma(t)$ to synthesize the transmit signal, i.e.,
\begin{equation}
    f_\text{Tx}(t) \coloneqq \sum_{({m},{n})\in \mathcal{I}} x_{m,n} \gamma(t-nT) \e^{2 \pi j mF t},
    \label{eq:tx}
\end{equation}
where 
\mbox{$\textcolor{black}{\vec x =} \{x_{m,n}\}_{(m,n) \in \mathcal I}$}
is the 2D-array of the \gls{tf} symbols containing data and pilot symbols.
The data symbols are modulated and encoded sequences of letters from a given alphabet generated by an information source. 
In contrast to the data symbols, the pilot symbols are known at the receiver and are coming from a different alphabet. 
\par
The doubly-dispersive channel model in the continuous \gls{dd} domain with a total of $R$ multipaths can be expressed as
\begin{equation}
    \eta(\tau, \nu) \coloneqq \sum_{r\in \mathcal J} \eta_{r} \delta(\tau-\tau_r)\delta(\nu-\nu_r),
    \label{eq:dd:eta}
\end{equation}
where 
the index set $\mathcal J = \{1, \dots, R\}$ associated with each path corresponds, respectively, to the delay shifts $\tau_{r}$, the Doppler shifts $\nu_r$, and the complex-valued attenuation factors $\eta_{r}$.
The assumption of the channel being underspread implies that all tuples $(\tau_{r}, \nu_r)$ are contained within a small region referred to as spreading region $\mathcal U \subset [0, \tau_\text{max}] \times [-\nu_\text{max}, \nu_\text{max}]$ such that 
$|\mathcal U|=2\tau_{\text{max}}\nu_{\text{max}}\ll 1$, where $\tau_{\text{max}}$ and $\nu_{\text{max}}$ correspond to the largest delay spread and largest Doppler spread, respectively  \cite{bello1963characterization}.
In the time domain, the channel in \eqref{eq:dd:eta} acts on the transmit signal in \eqref{eq:tx}  as a time-varying convolution. 
Hence the received signal yields
\begin{equation}
    f_\text{Rx}(t) \coloneqq \sum_{r \in \mathcal J} \eta_r f_\text{Tx}(t - \tau_r) \e^{2 \pi j\nu_r t}.
\end{equation}
The receiver analyzes the signal using another Gabor filterbank. 
We assume it uses the same Gabor grid as the transmitter and can only differ in the choice of the analysis pulse $g(t)$.
Then, we can describe the measured 2D-array of the \gls{tf} symbols  $\textcolor{black}{\vec y =} \{y_{\bar m, \bar n}\}_{(\bar m, \bar n) \in \mathcal I}$ by
%
\begin{multline}
        y_{\bar m,\bar n}
        =\int g^\ast(t-\bar n T) \e^{-2 \pi j \bar m F t}  f_\text{Rx}(t) \d t + w_{\bar m,\bar n},\\
        =\!\!\iiint \!\!g^\ast \!(t-\bar nT) \e^{j2 \pi t (\nu-\bar mF)} \eta(\tau, \nu)  f_\text{Tx}(t-\tau) d\tau d\nu dt+ w_{\bar m,\bar n},\\
        = \sum_{r\in\mathcal J} \eta_{r} \underbrace{\int g^\ast(t-\bar n T) \e^{2 \pi j t (\nu_r-\bar m F)} f_\text{Tx}(t-\tau_r) \d t}_{\eqqcolon y_{\bar m,\bar n}(\tau_r, \nu_r)}+w_{\bar m,\bar n},
        \label{eq-rx-full}
\end{multline}
where $\textcolor{black}{\vec y(\tau, \nu) =} \{y_{\bar m, \bar n}(\tau, \nu)\}_{(\bar m, \bar n) \in \mathcal I}$ is the 2D-array of the receiver response to a single unit amplitude scatterer where
$\tau$ is the delay shift, $\nu$ is the Doppler shift, and \mbox{$\textcolor{black}{\vec w=}\{w_{\bar{m},\bar{n}}\}_{(\bar{m},\bar{n})\in\mathcal{I}}$} is the 2D-array of the noise.
In our system model, we assume that the measured noise samples $w_{\bar{m},\bar{n}}$ are uncorrelated zero-mean random variables with variance $\sigma^{2}>0$.
The unit receiver response in \eqref{eq-rx-full} further evaluates to
 \begin{multline}
        y_{\bar m,\bar n}(\tau, \nu)
        = \int g^\ast(t-\bar nT) \e^{j2 \pi t (\nu-\bar mF)} 
        \\ \times \sum_{({m},{n})\in \mathcal{I}} x_{m,n} \gamma(t-\tau -nT) \e^{2 \pi j mF (t-\tau)} dt,\\
        = \sum_{({m},{n})\in \mathcal{I}} x_{m,n}\\
        \times \!\!\underbrace{\int g^\ast(t-\bar n T)\gamma(t-\tau-nT) \e^{2 \pi j(t\nu-\bar m Ft + mFt - mF \tau)} \d t}_{\eqqcolon \phi_{(m,n),(\bar m, \bar n)}(\tau,\nu)},
        \label{eq-rx-response}
 \end{multline}
where $\textcolor{black}{\vec \phi(\tau,\nu) =} \{\phi_{(m,n),(\bar m, \bar n)}(\tau,\nu)\}_{(m,n),(\bar m, \bar n) \in \mathcal I}$ is the effective channel matrix corresponding to a single unit amplitude scatterer.
It can be written as
%
\begin{equation}
    \begin{split}
         \phi_{(m,n),(\bar m, \bar n)}(\tau,\nu)=
         \e^{2 \pi j(\bar n T \nu-mF\tau +TF\bar n \Delta m)}\\ 
         \times \int  g^\ast(t)\gamma(t-\tau-\Delta n T)\e^{2 \pi j t(\nu+\Delta m F)}\d t,
        \label{poly}
    \end{split}
\end{equation}
where $\Delta n= n-\bar n$ and $\Delta m =  m -\bar m$ for convenience.
%
%
Observe that the integral in \eqref{poly} corresponds to the \emph{cross ambiguity function} of $\gamma$ and $g$ which we define as 
\begin{equation}
    A_{\gamma, g}(\tau, \nu) \coloneqq \int  g(t)^\ast \gamma(t-\tau) e^{2 \pi j \nu t} dt.
\end{equation}
The 2D-array of \gls{cmd} coefficients \mbox{$\textcolor{black}{\vec h(\tau,\nu) =} \{h_{\bar m,\bar n}(\tau,\nu)\}_{(\bar m,\bar n) \in \mathcal I}$} with respect to a single unit scatterer for $\Delta m=0$ and $\Delta n=0$ is given as
\begin{equation}
    h_{\bar m,\bar n}(\tau, \nu)
    \coloneqq 
      \phi_{(\bar m, \bar n),(\bar m, \bar n)}(\tau, \nu).
      \label{eq:cmd}
\end{equation}
\textcolor{black}{Due to the assumption of an underspread channel, the diagonal elements of the effective channel matrix are dominant \cite{kozek:nofdm,liu2004orthogonal,jung2007weyl,jung2007wssus}.}
This motivates the use of \gls{cmd} estimation which is significantly less complex than  maximum-likelihood estimation or iterative interference cancellation methods.
Exact orthogonality of the pulses in the integral of \eqref{poly} would imply that the effective channel matrix reduces to a diagonal matrix. 
This, however, cannot be achieved in pulse-shaped multicarrier systems independently of $(\tau, \nu)$ due to the intrinsic limitations of the \emph{cross ambiguity function} \cite{PfadlerWSA2020, PfadlerICC}.
To cope with this, we separate the off-diagonal terms in \eqref{poly} 
which cause the observed self-interference due to both inter-carrier and inter-symbol interference.
More specifically, we define the self-interference associated with a single unit amplitude scatterer as
\begin{equation}
    z_{\bar m,\bar  n}(\tau, \nu)
    \coloneqq \sum_{\substack{\mathclap{( m,n)\in \mathcal{I}}\\ (m,n) \neq (\bar m,\bar n)}}
        x_{m,n} \phi_{(m,n),(\bar m, \bar n)}(\tau,\nu) .
    \label{eq:z:scat}
\end{equation}
Then, we can write \eqref{eq-rx-response} with \eqref{eq:cmd} and \eqref{eq:z:scat} as 
\begin{equation}
    y_{\bar m,\bar n}(\tau, \nu)
    =
        x_{\bar m, \bar n} h_{\bar m,\bar n}(\tau, \nu)+z_{\bar m,\bar  n}(\tau, \nu),
    \label{eq-unit-response-approx}
\end{equation}
Finally, we can write the input-output relation in \eqref{eq-rx-full} with \eqref{eq-unit-response-approx} as
\begin{align}
    \vec y 
    &= \vec x \odot \sum_{r\in\mathcal J} \eta_{r} \vec h(\tau_r, \nu_r)  + \sum_{r\in\mathcal J} \eta_{r} \vec z(\tau_{r}, \nu_r) +\vec w,\\
    &= \vec x \odot \vec h + \vec z +\vec w,
    \label{eq-output-input-approx}
\end{align}
where 
\mbox{$\vec h = \{h_{\bar m, \bar n}\}_{(\bar m,\bar n) \in \mathcal I}$} and \mbox{$\vec z = \{z_{\bar m, \bar n}\}_{(\bar m,\bar n) \in \mathcal I}$} is the 2D-array of the \gls{cmd} and the 2D-array of self-interference, respectively.
We assume that the long-term expectation of the power over the normalized and zero-mean \gls{tf} symbols gives $\mathbb{E}\{|x_{m,n}|^2\}=1$.
Therefore, we model the distribution of $z_{\bar m, \bar n}$ as other random variables with uncorrelated zero mean noise and with (unknown) variance $\sigma_z^2 > 0$, which does not depend on $\bar{m}$ and $\bar{n}$ \cite{jung2007wssus,PfadlerICC}. 
\section{Linear Precoding and \gls{tf} Diversity \label{sec-spreading-diversity}}
We can significantly improve the performance of the multicarrier system by using a so-called linear precoding also referred to as spreading.
We point out that our model can be easily extended to include redundancy (i.e., number of rows is greater than number of columns of the precoding matrix) but, for the sake of simplicity, we restrict ourselves to orthogonal transforms.
Precoding and decoding are applied to the \gls{tf} symbols prior to Gabor synthesis and after Gabor analysis, respectively.
Generally, we refer to any energy-preserving linear mapping from information symbols $\vec X$ to \gls{tf} symbols $\vec x$ as linear precoding and its inversion as linear decoding, accordingly.
%
The key idea behind precoding is to intermingle information symbols such that each \gls{tf} symbol contains information on all information symbols, which turns out to enable high \gls{tf} diversity at the receiver \cite{zemen2018iterative,tharaj2021otsm}. 
The precoding distributes equalization errors and self-interference evenly across all information symbols, so that per symbol modulation works more reliable.
This is important since -- for example -- the equalization error becomes locally large near zero-crossings of the \gls{cmd} coefficients.
In turn, the \gls{ber} is significantly reduced.

\gls{otfs} is a notable example that applies jointly orthogonal precoding to both data and pilot symbols.
In \gls{otfs}, all symbols \mbox{$\vec X = \{X_{\ell, k}\}_{(\ell, k) \in \mathcal{I}^\circ}$} are placed in the \gls{dd} domain and then transformed into the \gls{tf} domain by applying the \gls{2D-dsft}, i.e.,
\begin{equation}
    x_{m,n} 
    = 
    \frac{1}{\sqrt{NM}} 
    \sum_{(\ell,k) \in \mathcal{I}^\circ} X_{\ell,k} e^{-j2 \pi (\frac{n \ell}{N}-\frac{m k}{M})}, 
  \label{eq-dsft}
\end{equation}
where we use $\mathcal I^\circ = \Z_N \times \Z_M$ as indices of the adjoint grid $\Lambda ^\circ = T^{-1} \Z_N \times F^{-1} \Z_M$ corresponding to the \gls{dd} domain.
The \gls{2D-dsft} in \cref{eq-dsft} is its own inverse as a result of opposite exponential sign, the flipping of the axes, and normalization; hence, orthogonal precoding and orthogonal decoding are the same operation.\footnote{LTFAT \url{http://ltfat.org/doc/gabor/dsft.html}} 
To some extent, the choice of the \gls{2D-dsft} for orthogonal precoding is motivated by
\cref{eq-rx-response,eq-output-input-approx}, which show that 
\begin{multline}
    (\bar m, \bar n) \mapsto h_{\bar m, \bar n} = \sum_{r \in \mathcal J} 
        \eta_{r}  
        \e^{2 \pi j(\bar n T \nu_r -\bar mF\tau_r)}\\
        \times \int g^\ast(t)\gamma(t-\tau_r)\e^{2 \pi j t\nu_r}\d t
        \label{eq:trigo:polynomial}
\end{multline}
 are the samples of a low-frequency 2D trigonometric polynomial which corresponds to Dirac pulses in the continuous \gls{dd} domain.
Many \gls{otfs} channel estimation schemes aim at making use of this fact, and it has been topic of many research to harness the sparsity of the channel \cite{Zhang2018, shen2019}.
Since our proposed channel estimation scheme only precodes data symbols, the orthogonal basis function is independent of the proposed piloting scheme and the choice of it is arbitrary
as long as maximum \gls{tf} diversity is achieved.
\section{Channel Estimation and Equalization}
\label{section-channel-estimator}
In this section, we discuss the estimation of doubly-dispersive channels and leakage effects. 
We present the proposed channel estimator using smoothness optimization and detail its design choice as well as pilot signaling, equalization, and data recovery.
\subsection{Leakage effects \label{section-leakage}}
In pulse-shaped multicarrier systems, the sparsity of the channel diminishes after applying discrete Fourier transforms 
to the received symbols after the Gabor analysis filterbank.
To see this, we compute the 2D-array of the \gls{cmd} for a $(\tau,\nu)$-scatterer in \gls{dd} domain by applying the \gls{2D-dsft} to $ h_{\bar m,\bar n}(\tau, \nu)$ in \eqref{eq:cmd} as \cite{PfadlerWCL}
%
\begin{multline}
    H_{\bar \ell, \bar k}(\tau, \nu)
    = \sum_{(\bar m,\bar n)\in \mathcal{I}} 
        h_{\bar m,\bar n}(\tau, \nu) \, \e^{-2\pi j (\frac{\bar m \bar k}{M} - \frac{\bar n \bar \ell}{N})},\\
    = A_{\gamma, g}(\tau, \nu) 
        \sum_{\bar n = 0}^{N-1} \e^{2 \pi j \frac{\bar n (\bar \ell + NT\nu)}{N}}
        \sum_{\bar m=0}^{M-1} \e^{-2 \pi j \frac{\bar m (\bar k + MF \tau)}{M}},\\
    = A_{\gamma, g}(\tau, \nu) 
        D_N \left( \frac{\bar \ell +  NT\nu}{N} \right)
        {D_M \left(\frac{-\bar k -  MF\tau}{M} \right)},
    \label{eq:leakage:3}
\end{multline}
where $D_K$ corresponds to the \textit{Dirichlet kernel} for an integer $K > 0$ which is defined to be
\begin{equation}
    D_K(t)
    \coloneqq \sum_{k=0}^{K-1} \e^{2\pi j k t}
    = \begin{cases}
        K, & \text{if } t \in \Z,\\
       \e^{\pi j (K-1) t} \frac{\sin(\pi K t)}{\sin(\pi t)}, & \text{otherwise.} 
    \end{cases}
\end{equation}
The 2D-array \mbox{$\textcolor{black}{\vec H(\tau, \nu) =} \{H_{\bar \ell,\bar k}(\tau, \nu)\}_{(\bar \ell,\bar k) \in \mathcal I^\circ}$} is only sparse if both $NT \nu$ and $ MF \tau$ are integers.
This is exactly not the case when fractional delay shifts and fractional Doppler shifts are present, thereby causing the observed leakage effects.
Moreover, discrete Fourier transforms assume that their input samples stem from a bandlimited and periodic function, which is not true in our setup.
The observed leakage of an individual scatterer follows the shape of the poorly localized Dirichlet kernel for the most part,
whereas the design of the \emph{cross ambiguity function} has a comparably marginal impact on it, see also \cite{PfadlerWCL}. 
This degrades the performance of the channel estimation significantly unless more expensive leakage suppression techniques are applied, cf.~\cite{wei2021window, PfadlerWCL}.
\subsection{General piloting scheme}
\label{section-proposed-piloting}
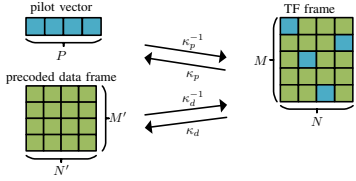
\begin{figure}[t]
  \centering 
  \scalebox{0.5}{
  \begin{tikzpicture}[]
    \node at (0cm,0cm) {\includegraphics[width=0.49\textwidth]{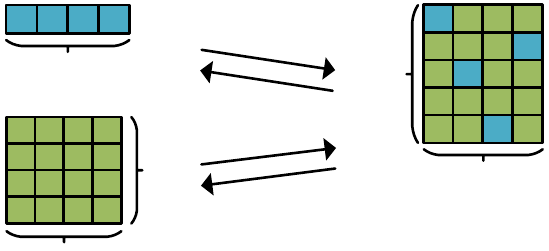}};
    \node at (0.1cm,1.3cm)[rotate=0,align=center]{ $\kappa_p^{-1}$}; 
    \node at (0.1cm,0.4cm)[rotate=0,align=center]{ $\kappa_p$}; 
    \node at (0.1cm,-0.2cm)[rotate=0,align=center]{ $\kappa_d^{-1}$}; 
    \node at (0.1cm,-1.1cm)[rotate=0,align=center]{ $\kappa_d$}; 
    \node at (3.4cm,-0.8cm)[rotate=0,align=center]{ $N$}; 
    \node at (1.9cm,0.8cm)[rotate=0,align=center]{ $M$}; 
    \node at (-3.4cm,-2.1cm)[rotate=0,align=center]{ $N'$}; 
    \node at (-1.9cm,-0.7cm)[rotate=0,align=center]{ $M'$}; 
    \node at (-3.4cm, 0.95cm)[rotate=0,align=center]{ $P$}; 
    \node at (3.2cm,2.2cm)[rotate=0,align=center]{ {TF frame}}; 
    \node at (-3.4cm,0.35cm)[rotate=0,align=center]{ {precoded data frame}}; 
    \node at (-3.4cm,2.2cm)[rotate=0,align=center]{ {pilot vector}}; 
  \end{tikzpicture}}
  \caption{Multi- and demultiplexing of the pilot vector and precoded data frame in the \gls{tf} domain.}
  \label{fig:mapping}  
\end{figure}

Our goal is to estimate the \gls{cmd} coefficients $\vec h$ from the pilot symbols of the received frame $\vec y$ in \eqref{eq-output-input-approx}. 
Motivated by the 2D orthogonal precoding via the \gls{2D-dsft}, we let the data symbols originate from a 2D data frame indexed by a Gabor grid $\mathcal I' = \Z_{M'} \times \Z_{N'}$ with $M' \leq M$ and $N' \leq N$.
Then, the symbols from the data frame are multiplexed with the pilots into the \gls{tf} frame.
Therefore, we define the index set of the data symbols in the \gls{tf} frame as
$\mathcal D \subset \mathcal I$ satisfying $\# \mathcal D = \#\mathcal I' = M'N'$. 
For the indices of the pilot symbols, we take the complement set $\mathcal P = \mathcal I \setminus \mathcal D$ and put $P = \# \mathcal P$ as the number of pilots.
We choose arbitrary bijective maps $\kappa_d : \mathcal D \to \mathcal I'$ and $\kappa_p : \mathcal P \to \{1, \dots, P\}$ which describe how the data and pilot symbols are mapped onto the transmitted and received \gls{tf} frame.
We illustrate this mapping in Fig. \ref{fig:mapping}.  

Given a fully precoded 2D-array 
$\vec x' = \{x'_{m', n'}\}_{(m', n') \in \mathcal I'}$ of data symbols
and a vector of pilot symbols $\vec p \in \C^{P}$,
we define the content of the \gls{tf} frame by the following multiplexing:
\begin{equation}
    x_{m,n}
    = \begin{cases}
        x'_{\kappa_d(m, n)}, & \text{if } (m, n) \in \mathcal D,\\
        p_{\kappa_p(m, n)}, & \text{if } (m, n) \in \mathcal P.
    \end{cases}
    \label{eq-tf-piloting}
\end{equation}
The ordering of the elements $\vec x'$ does not impact the achievable \gls{tf} diversity gain when using orthogonal precoding transformations \cite{zemen2018iterative}.
For this reason the choice of $\kappa_d$ and $\kappa_p$ does not impact the performance, whereas the size of $\mathcal D$ and $\mathcal P$ does.

At the receiver, we can then extract the distorted pilot vector $\vec q \in \C^{P\times 1}$ from the received frame in \eqref{eq-output-input-approx} by
\begin{equation}
      q_{s} =  y_{\kappa_p^{-1}(s)} \qquad s = 1, \dots, P.
      \label{eq:pilot-rx}
\end{equation}
Then, we can estimate the channel from \eqref{eq:pilot-rx} by different schemes as described in the following subsections. 

\subsection{Standard \gls{lmmse} estimator \label{sub:lmmse}}


The \gls{lmmse} estimator, which is a \gls{dft}-based estimator, follows a regularized \gls{ls} scheme as discussed in \cite{hoeher1997two,luenberger1997optimization,jung2009robust}. 
This scheme assumes that most of the energy of \gls{cmd} coefficients is concentrated near the origin in the \gls{dd} domain.
The least-squares reconstruction is then performed on a subset of the DD domain which we refer to as reconstruction grid. 
We reduce the degrees of freedom by enforcing the estimated \gls{cmd} to be zero outside the reconstruction grid, which we define as 
%
    \begin{equation}
        \mathcal{K}= \{-Q, \dots, Q\} \times \{-W_{\text{n}}, \dots, W\} \subset\mathcal{I}^\circ,
        \label{eq:reconstruction:grid}
    \end{equation}
where $Q$ and $W$ specify the reconstruction grid for the expected shifts in Doppler domain and delay domain, respectively.
They need to be selected such that $Q > \nu_\text{max}TN$ and $W >\tau_\text{max}FM$.
However, due to the poor resolution of the Dirichlet kernel in \eqref{eq:leakage:3}, fractional shifts are smeared over the \gls{dd} grid. 
As a consequence, the reconstruction grid has to be expanded.
In the case of smeared Doppler shifts, we just increase $Q$ since they are generally distributed symmetrically to the origin.
In contrast, delay shifts are distributed asymmetrically. 
Therefore, we introduce the parameter $W_{\text{n}}$ to consider smeared delays close to the origin.
\par
We start with the initial partial \gls{cmd} estimate $\vec h^\text{pilot}\in \C^{P}$ given as  
\begin{equation}
    h^\text{pilot}_{s} =  q_{s} /  p_{s} \qquad s = 1, \dots, P.
    \label{eq-partial-cmde}
\end{equation}
To complete the \gls{cmd} estimate, we search for the best \gls{ls} fit among all \gls{cmd}s which are supported on the reconstruction grid. 
For this, we define a sub-matrix $\vec C \in \C^{P \times 2Q(W+W_\text{n})}$ of the \gls{2D-dsft} to link the reconstruction grid to the pilot symbols in \gls{tf} domain as
\begin{equation}
     C_{(\bar m,\bar n),(\bar \ell,\bar k)} 
    = \frac{1}{\sqrt{NM}} e^{-j2 \pi (\frac{\bar n \bar \ell}{N}-\frac{\bar m \bar k}{M})}, 
  \label{eq-ls-est}
\end{equation}
where $ (\bar m, \bar n) \in \mathcal P \,\, \text{and} \,\,(\bar \ell, \bar k) \in \mathcal K$.
With this in hand, we can formulate the optimization problem as
\begin{equation}
    \operatornamewithlimits{min}_{\vec{\tilde{H}}}
         \|\vec h^\text{pilot}- \vec C \vectorize \{\tilde{ \vec H}\}\|^2_2 , 
    \label{proposed-problem}
\end{equation}
where 
\mbox{$\vec{ \tilde{H}} = \{{ \tilde{H}}_{\bar \ell,\bar k}\}_{(\bar \ell,\bar k) \in \mathcal{K}}$} is the 2D-array of the estimated \gls{cmd}.
The optimization problem in \eqref{proposed-problem}, has a closed form solution which is given by
\begin{equation}
    \tilde{\vec{H}}^{\text{LMMSE}}
        = (\vec C^\text{H}\vec C + \vec I\sigma^2)^{-1}\vec C^\text{H}\vec h^\text{pilot},
    \label{eq-ls}
\end{equation}
where $\vec I$ is the $ 2Q(W+W_\text{n}) \times 2Q(W+W_\text{n})$ identity matrix. 
Finally, we transform the estimated \gls{cmd} coefficients of \eqref{eq-ls} to the \gls{tf} domain
by applying a \gls{2D-dsft}, i.e., 
\begin{equation}
    \tilde{{h}}^{\text{LMMSE}}_{\bar m,\bar n} 
    = 
    \frac{1}{\sqrt{NM}} 
    \sum_{(\bar \ell,\bar k) \in \mathcal{K}} \tilde{{H}}^{\text{LMMSE}}_{\bar \ell,\bar k} e^{-j2 \pi (\frac{\bar n \bar \ell}{N}-\frac{\bar m \bar k}{M})}. 
  \label{eq-dsft-lmmse}
\end{equation}

\subsection{Proposed smoothness regularized channel estimator \label{sub:proposed:est}}
We propose to estimate the \gls{cmd} coefficients in the \gls{tf} domain to avoid leakage observed in the discrete \gls{dd} domain.
Our scheme estimates the channel by interpolating smooth functions from the received pilot symbols.
This is achieved by a novel regularizer which minimizes the energy of the second order derivatives.
To justify this, we point out that in \eqref{eq:trigo:polynomial} the channel in the continuous \gls{dd} domain consists of samples which are 2D trigonometric polynomials and low-frequency meaning that they are relatively slow changing compared to the frame size. 
In general, it is known that the second order derivative is a measure for the smoothness of functions. 
To smooth such functions, it is a common approach to minimize the second order derivative of the samples. 
The proposed channel estimation scheme follows this approach.
\par
We compute the second order discrete derivatives using non-cyclic convolutions with kernels of the size $3 \times 3$.
Specifically in our setup, the 2D convolution of an array $\vec E = \{E_{\bar m,\bar n}\}$ of size $(M+2) \times (N+2)$ with a $3 \times 3$ kernel $\vec \Phi = \{\Phi_{\bar m,\bar n}\}$ is the array
\begin{align}
    [\vec E \ast \vec \Phi]_{\bar m,\bar n} 
    = \sum_{\bar \ell = -1}^{1} \sum_{\bar k = -1}^{1} E_{\bar m-\bar \ell+1, \bar n-\bar k+1} \Phi_{\bar \ell+1, \bar k+1},\\
    \quad \bar m = 0, \ldots, M-1,\quad \bar n = 0, \ldots N-1,
\end{align}
of size $M \times N$, i.e., we consider the valid part of the convolution.
We define the kernels as 
\begin{equation}
    \begin{split}
        \vec \Phi_{\text{tt}}
        &= \begin{bmatrix} \phantom{-}0 & \phantom{-}0 & \phantom{-}0\phantom{-} \\ -1 & \phantom{-}2 & -1\phantom{-} \\ \phantom{-}0 & \phantom{-}0 & \phantom{-}0\phantom{-} \end{bmatrix},
        \quad 
        \vec \Phi_{\text{ff}} 
        = \begin{bmatrix} \phantom{-}0 & -1 & \phantom{-}0\phantom{-} \\ \phantom{-}0 & \phantom{-}2 & \phantom{-}0\phantom{-} \\ \phantom{-}0 & -1 & \phantom{-}0\phantom{-} \end{bmatrix},
        \\
        \vec \Phi_{\text{tf}}
        &= \begin{bmatrix} -1 & \phantom{-}1 & \phantom{-}0\phantom{-} \\ \phantom{-}1 & -1 & \phantom{-}0\phantom{-} \\ \phantom{-}0 & \phantom{-}0 & \phantom{-}0\phantom{-} \end{bmatrix},
    \end{split}
\end{equation}
where $\vec \Phi_{\text{ff}}, \vec \Phi_{\text{tt}}$ and $\vec \Phi_{\text{tf}}$ correspond to the second order partial derivatives with respect to frequency, time and mixed dimensions, respectively.
With this in hand, we define the discrete weighted Hessian with $\bar m = 0, \ldots, M-1, \bar n = 0, \ldots, N-1$ as
\begin{equation}
    \label{eq-weighted-discrete-hessian}
    \vec Q^{\alpha, \beta}_{\vec E}(\bar m, \bar n)
    = \begin{bmatrix} 
        \alpha^2 \vec [\vec E \ast \vec \Phi_{\text{ff}}]_{\bar m, \bar n}
        & \alpha \beta [\vec E \ast \vec\Phi_{\text{tf}}]_{\bar m, \bar n}\\
        \alpha \beta [\vec E \ast \vec\Phi_{\text{tf}}]_{\bar m, \bar n}
        & \beta^2 [\vec E \ast \vec\Phi_{\text{tt}}]_{\bar m, \bar n}
    \end{bmatrix},
\end{equation}
where the scaling parameters $\alpha, \beta > 0$ assist in compensating channel modes which we detail in \cref{section-scaling}.

The proposed channel estimator provides a solution to the optimization problem given as
\begin{equation}
    \begin{split}
    &\operatornamewithlimits{min}_{\vec h^\text{ex}}
    \sum_{\bar m=0}^{M-1}\sum_{\bar n=0}^{N-1} 
        \| \vec Q^{\alpha, \beta}_{\vec h^\text{ex}}(\bar m, \bar n) \|_{\text{F}}^2 \\
    &\operatorname{subject\ to} \quad \begin{cases}
        {\vec h^\text{ex}} \in \C^{(M+2) \times (N+2)}, \\
        \sum_{s = 1}^P | h^\text{pilot}_{s} - h^\text{ex}_{\kappa_p^{-1}(s)} |^2 \leq \delta,
    \end{cases}
    \label{eq-proposed-channel-estim}
    \end{split}
\end{equation}
where $\delta$ is a relaxation parameter and $h^\text{ex}$ is an array containing the \gls{cmd} estimate with appropriate padding to compensate for the size reduction from the convolution. 
The optimization problem in \eqref{eq-proposed-channel-estim} is a convex constrained \gls{ls} problem which can effectively be solved by standard methods \cite{boyd2004convex}.
The actual \gls{cmd} estimate $\tilde {\vec h}$ is then obtained by truncating $\vec h^\text{ex}$ at the frame boundaries, i.e., 
\begin{equation}
    \tilde h_{\bar m, \bar n}
    = h^\text{ex}_{\bar m, \bar n}, \quad \bar m = 1, \ldots, M, \bar n = 1, \ldots N.
\end{equation}

\subsection{Awareness of the channel mode}
\label{section-scaling}
The scaling factors $\alpha$ and $\beta$ in \eqref{eq-proposed-channel-estim} control the preferred \emph{channel mode} for the reconstruction, defined below.
Let us briefly explain the intuition behind the weighting.
The 2D-array of the \gls{cmd} coefficients is in fact a sampling of an underlying differentiable function $h(f, t)$ as shown in \eqref{eq:trigo:polynomial}.
In essence, the \emph{mode} of a channel is given by the ratio of its 2D-support in the \gls{dd} domain. 
Suppose $h(f,t)$ is approximately supported on the rectangular box \mbox{$[-\beta, \beta] \times [-\alpha, \alpha]$} in \gls{dd} domain.
Writing \mbox{$h(\alpha f, \beta t) = u(f,  t)$}, we have that $u$ in \gls{dd} domain is supported on the unit square \mbox{$[-1,1] \times [-1, 1]$} and its mode is balanced between delay domain and Doppler domain.
In \cref{eq-proposed-channel-estim}, it is beneficial to regularize on the Hessian of $u$ rather than $h$ as the regularizing term does not favor any particular direction.
By standard calculus, we know that the (continuous) Hessian matrix of $u$ at the point $(f,t) \in \R^2$ is given by 
\begin{equation}
    \begin{bmatrix} 
        \alpha^{2} \frac{\partial^2}{\partial f^2} h(\alpha f, \beta t) & 
        \alpha \beta \frac{\partial^2}{\partial f \partial t} h(\alpha f, \beta t)\\
        \alpha \beta \frac{\partial^2}{\partial f \partial t} h(\alpha f, \beta t) & 
        \beta^{2} \frac{\partial^2}{\partial t^2} h(\alpha f, \beta t)
    \end{bmatrix}.
\end{equation}
As we only have access to the (discrete) Hessian of $\vec h$, we include additional scaling into the optimization manually, obtaining the weighted discrete Hessian matrix as in \cref{eq-weighted-discrete-hessian}. 

In summary, given that the doubly-dispersive channel in \eqref{eq:dd:eta} has maximum delay spread $\tau_\text{max}$ and maximum Doppler spread $\nu_\text{max}$, a reasonable choice is to put $\alpha = \nu_\text{max}$ and $\beta = \tau_\text{max}$.
However, depending on other factors, for example, if the contribution of many scatterers is negligible, the parameters should be adjusted accordingly.
\subsection{Noise-awareness}
\label{section-noise}
We relaxed the data fidelity term in \cref{eq-proposed-channel-estim} to mitigate noisy measurements.
The relaxation parameter $\delta$ needs to match the expected error given as
\begin{equation}
    \delta = \mathbb E \bigl\{\sum_{s = 1}^P 
        | h_{s}^\text{pilot} - h_{\kappa_p^{-1}(s)} |^2 \bigr\}.
\end{equation}
Considering \eqref{eq-partial-cmde} with noise and self-interference, we get the initial \gls{cmd} estimation as
\begin{equation}
     h^\text{pilot}_{s} 
    = \frac{q_{s}}{p_{s}}
    = h_{\kappa_p^{-1}(s)} + \frac{z_{\kappa_p^{-1}(s)}}{p_{s}}+\frac{w_{\kappa_p^{-1}(s)}}{p_{s}}
\end{equation}
and thus 
\begin{equation}
    \begin{split}
        \mathbb E \bigl\{ | h_{s}^\text{pilot} - h_{\kappa_p^{-1}(s)} |^2 \bigr\}
        &= \frac{\mathbb E \bigl\{|z_{\kappa_p^{-1}(s)}|^2\bigr\}}{|p_{s}|^2}+\frac{\mathbb E \bigl\{|w_{\kappa_p^{-1}(s)}|^2\bigr\}}{|p_{s}|^2},\\
        &= (\sigma_z^2 +\sigma^2) |p_{s}|^{-2}.
    \end{split}
\end{equation}
Hence, we choose the relaxation parameter as
\begin{equation}
    \delta = (\sigma_z^2 +\sigma^2)\sum_{s = 1}^P |p_{s}|^{-2}.
    \label{eq:relaxation:par}
\end{equation}
We can simplify \eqref{eq:relaxation:par} to \mbox{$\delta = (\sigma^2+\sigma_z^2) P$}, by considering the pilots to be normalized to unit energy per symbols, i.e., $\mathbb{E}\{|p_s|^2\}=1$.
\subsection{Pilot placement}
\label{section-pilot-placment}
Most relevant to the performance of the proposed channel estimation scheme is the choice of pilot positions, represented by the set $\mathcal P$.
As we optimize second order derivatives, we have to be aware that 
the approximation error in
$\vec h^\text{ex}$ tends to grow quadratically in the distance to the nearest pilot.
For that reason, it is best if $\mathcal P$ is distributed as uniformly as possible within $\mathcal I$.
This matter is complicated by the fact that orthogonal precoding transformations, such as the \gls{dsft} or \gls{fwht}, work best if $M'$ and $N'$ are powers of $2$.
We are therefore targeting a transmit frame size of $M \times N$ and have $P = NM - N'M'$ pilots. 
It is however not obvious how to distribute these uniformly in general.
To remedy this, we propose a piloting scheme 
in \cref{section-accordion}.

\subsection{Complexity of estimators}
\label{section-complexity}
Let us discuss some complexity aspects of the considered optimization problems. 
For the standard \gls{lmmse} estimator, we need to solve the unconstrained linear \gls{ls} problem in \eqref{proposed-problem}. 
The complexity of \eqref{proposed-problem} usually grows cubically with the frame size, i.e., by $(NM)^3$. 
However, due to the closed form solution in \eqref{eq-ls} the \gls{ls} estimator matrix can be computed for each $\sigma^2$ offline.
Then, the \gls{ls} problem is reduced to a matrix vector product for a fixed $\sigma^2$.
Regarding the complexity of the proposed estimator scheme, we need to solve the optimization problem in \cref{eq-proposed-channel-estim}.
The problem in \cref{eq-proposed-channel-estim} is however a constrained \gls{ls} problem and does not have a closed form solution.
By using Tikhonov regularization \cite{golub1999tikhonov}, we can convert the constraint problem in \cref{eq-proposed-channel-estim} into an unconstrained \gls{ls} problems as 
\begin{equation}
    \begin{split}
    &\operatornamewithlimits{min}_{\vec h^\text{ex}}
    \sum_{\bar m=0}^{M-1}\sum_{\bar n=0}^{N-1} 
        \| \vec Q^{\alpha, \beta}_{\vec h^\text{ex}}(\bar m, \bar n) \|_F^2 + \Omega \sum_{s = 1}^P | h^\text{pilot}_{s} - h^\text{ex}_{\kappa_p^{-1}(s)} |^2,
    \label{eq:proposed-channel-d2}
    \end{split}
\end{equation}

where $\Omega$ is another regularization parameter. 
Then, for each $\Omega$ a closed form solution exists which can be computed offline as for the \gls{lmmse} estimator. 

\subsection{Other considerations of the proposed estimator}
\label{section-other}
Let us discuss some other design choices and aspects of the proposed channel estimator in \cref{eq-proposed-channel-estim}.
Our first design decision is to extend the optimization variable $\vec h^\text{ex}$ rather than using padded convolution.
Regarding the most common padding techniques for convolution, we observe that:
\begin{itemize}
    \item
    Zero-padding causes a significant amplitude drop near the frame boundaries, which does not fit our model for $\vec h$ as seen in \cref{eq-output-input-approx}.
    \item
    Mirror-padding favors solutions that are flat at the frame boundaries.
    Although performing better than zero-padding, it still yields inferior estimations compared to the extension approach.
    \item
    Circular padding leads to a leakage effect similarly to the \gls{otfs} piloting scheme in \gls{dd} domain, cf.~\cref{section-leakage}.
\end{itemize}
\vspace{0.2em}
Let us explain the specific choice of minimizing the energy of the second order derivative.
Minimizing the gradient does not yield satisfactory results, as the trigonometric polynomials making up the true solution are not close to being (piece-wise) linear.
Using higher order derivatives requires more and larger kernels and therefore more computational time and memory.
In addition, the derivatives are less stable and often cause unreasonably large values to appear in the solution, especially near the frame boundaries.
In fact, we found no significant improvements in performance for third order derivatives and even worse performance for derivatives of greater order.


\subsection{Equalization and data recovery}
\label{section-equalization}
Let us detail the equalizer to construct the transmitted \gls{tf} frame from the received \gls{tf} frame with the estimated channel and the recovery of the transmitted bits.
The choice of a suitable equalization should be made based on the selected channel estimation scheme.
Recall that we estimate the \gls{cmd} and not the effective channel matrix with off-diagonal terms.
We furthermore aim at a data recovery on a per frame basis not considering iterative schemes. 
This makes one-tap equalization to a suitable scheme which we follow in this paper. 
We use a linear \gls{mmse} equalizer and get the equalized \gls{tf} frame as  
\begin{equation}
    \hat x_{\bar m,\bar n} = \tilde h_{\bar m,\bar n}^\ast y_{\bar m,\bar n}(| \tilde h_{\bar m,\bar n}|^2 + \sigma^2)^{-1}.
    \label{eq:mmse}
\end{equation}
We demultiplex the \gls{tf} frame to extract the precoded data frame $\hat{\vec x}' \in \C^{M'N'}$ by
\begin{equation}
     \hat x'_{\bar m', \bar n'} 
     = \hat x_{\kappa_d^{-1}(\bar m', \bar n')} \qquad (\bar m', \bar n') \in \mathcal I'.
\end{equation}
Then, $\hat{\vec x}'$ is linearly decoded, demodulated and decoded, yielding the transmitted information bits.
%
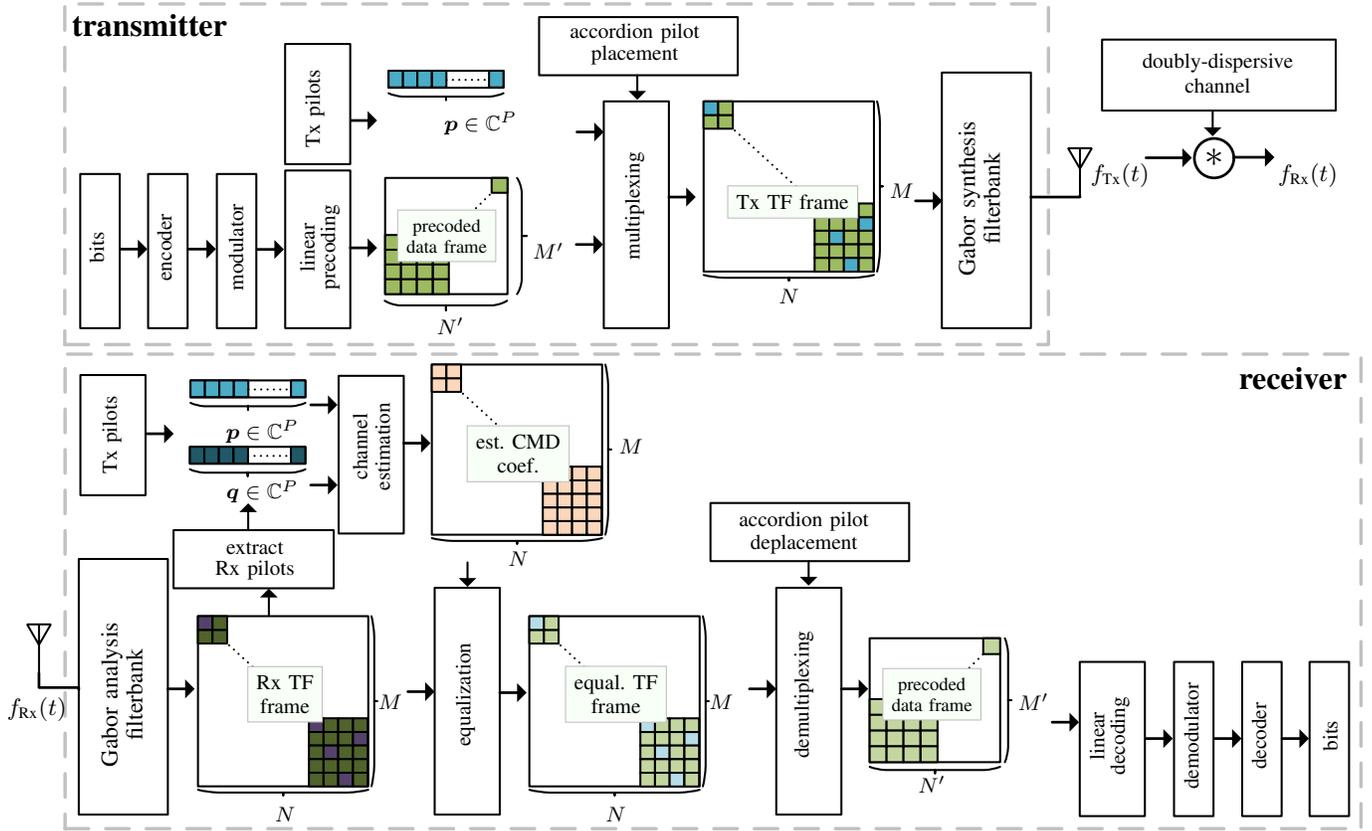
\begin{figure*}[t]
  \centering 
  \scalebox{1.0}{
  \begin{tikzpicture}[]
    \node at (0cm,0cm) {\includegraphics[width=0.99\textwidth]{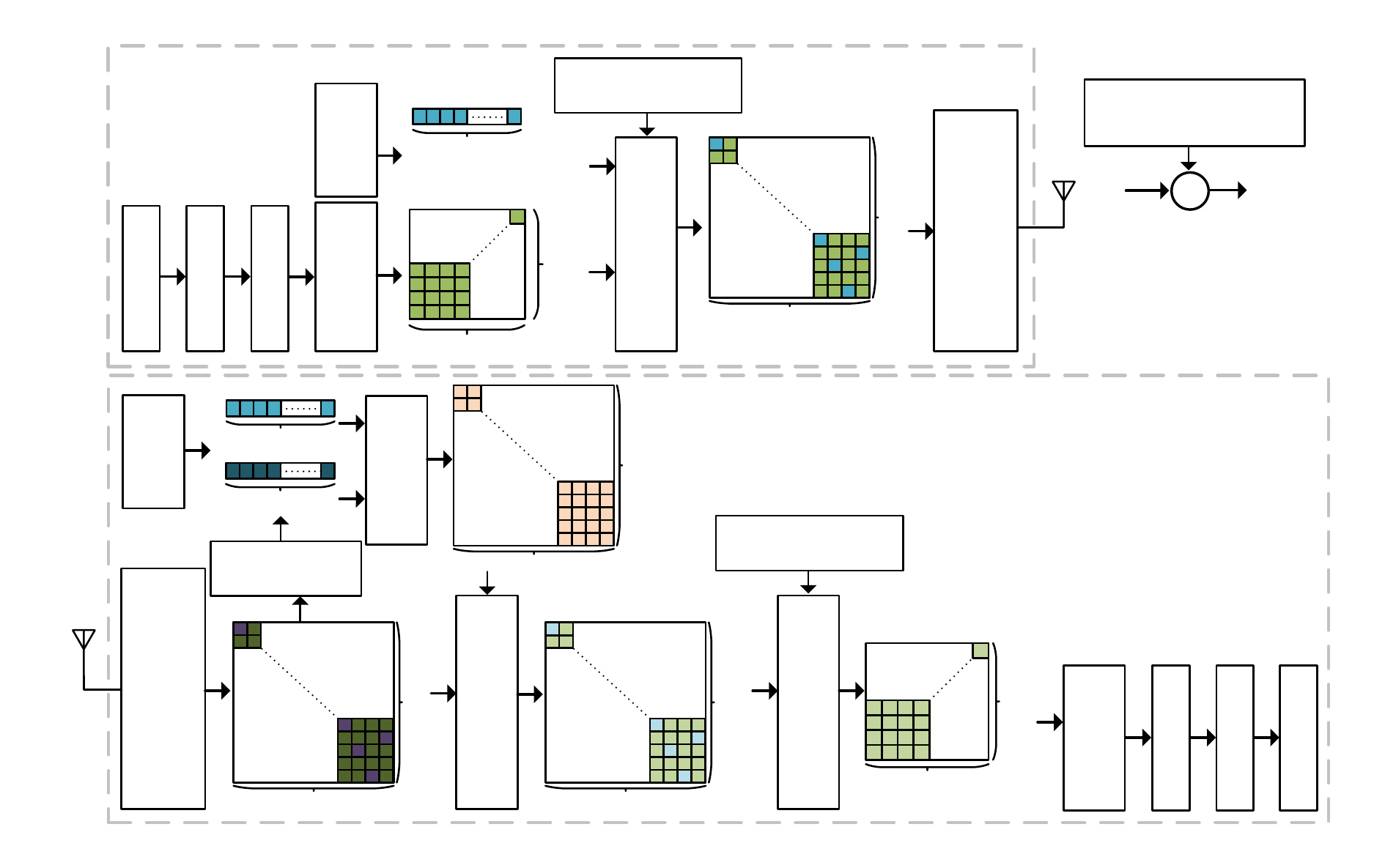}};
    \node at (-7.3cm,5.2cm)[rotate=0,align=center]{\large \textbf{transmitter}}; 
    \node at (-8.0cm,2.25cm)[rotate=90,align=center]{\footnotesize bits};
    \node at (-7.1cm,2.25cm)[rotate=90,align=center]{\footnotesize encoder}; 
    \node at (-6.15cm,2.25cm)[rotate=90,align=center]{\footnotesize modulator}; 
    \node at (-5.2cm,2.25cm)[rotate=90,align=center]{\footnotesize linear}; 
    \node at (-4.9cm,2.25cm)[rotate=90,align=center]{\footnotesize precoding}; 
    \node at (-5.1cm,4.1cm)[rotate=90,align=center]{\footnotesize Tx pilots}; 
    \node at (-0.85cm,2.8cm)[rotate=90,align=center]{\footnotesize multiplexing};
    \node at (-0.85cm,5.1cm)[rotate=0,align=center]{\footnotesize accordion pilot}; 
    \node at (-0.85cm,4.8cm)[rotate=0,align=center]{\footnotesize placement}; 
    \node at (6.9cm,4.7cm)[rotate=0,align=center]{\footnotesize doubly-dispersive}; 
    \node at (6.9cm,4.4cm)[rotate=0,align=center]{\footnotesize channel};
    \node at (6.85cm,3.45cm)[rotate=0,align=center]{\Large $\ast$}; 
    \node at (3.6cm,2.9cm)[rotate=90,align=center]{\small Gabor synthesis};
    \node at (3.9cm,2.9cm)[rotate=90,align=center]{\small filterbank}; 
    \node at (-2.9cm,3.9cm)[rotate=0,align=center]{\footnotesize $\vec p \in \C^{P}$};
    \node at (-2.0cm,2.2cm)[rotate=0,align=center]{\footnotesize $M'$};
    \node at (-3.3cm,1.2cm)[rotate=0,align=center]{\footnotesize $N'$};
    \node at (2.7cm,3.0cm)[rotate=0,align=center]{\footnotesize $M$};
    \node at (1.2cm,1.6cm)[rotate=0,align=center]{\footnotesize $N$};
    \node at (8.1cm,3.2cm)[rotate=0,align=center]{\small $f_\text{Rx}(t)$};
    \node at (5.6cm,3.2cm)[rotate=0,align=center]{\small $f_\text{Tx}(t)$};
    \node[rectangle,draw = lightgray,text = black,fill = green!3] (r) at (1.25cm, 2.85cm) {\footnotesize Tx TF frame};
    \node[rectangle,draw = lightgray,text = black,fill = green!3] (r) at (-3.35cm, 2.4cm) {\scriptsize \Umbruch{precoded data frame}};
    \node at (7.9cm,0.5cm)[rotate=0,align=center]{\large \textbf{receiver}}; 
    \node at (8.4cm,-4.25cm)[rotate=90,align=center]{\footnotesize bits}; 
    \node at (7.45cm,-4.25cm)[rotate=90,align=center]{\footnotesize decoder};
    \node at (6.55cm,-4.25cm)[rotate=90,align=center]{\footnotesize demodulator};
    \node at (-7.8cm,-0.25cm)[rotate=90,align=center]{\footnotesize Tx pilots}; 
    \node at (-5.8cm,-1.0cm)[rotate=0,align=center]{\footnotesize $\vec q \in \C^{P}$};
    \node at (-5.8cm,-0.2cm)[rotate=0,align=center]{\footnotesize $\vec p \in \C^{P}$};
    \node at (-7.8cm,-3.7cm)[rotate=90,align=center]{\small Gabor analysis};
    \node at (-7.5cm,-3.7cm)[rotate=90,align=center]{\small filterbank}; 
    \node at (-8.8cm,-3.9cm)[rotate=0,align=center]{\small $f_\text{Rx}(t)$};
    \node at (-3.1cm,-3.7cm)[rotate=90,align=center]{\footnotesize equalization};
    \node at (-4.5cm,-0.4cm)[rotate=90,align=center]{\footnotesize channel};
    \node at (-4.2cm,-0.4cm)[rotate=90,align=center]{\footnotesize estimation};
    \node at (1.4cm,-3.7cm)[rotate=90,align=center]{\footnotesize demultiplexing};
    \node at (5.3cm,-4.25cm)[rotate=90,align=center]{\footnotesize linear};
    \node at (5.6cm,-4.25cm)[rotate=90,align=center]{\footnotesize decoding};
    \node at (1.4cm,-1.4cm)[rotate=0,align=center]{\footnotesize accordion pilot};    
    \node at (1.4cm,-1.7cm)[rotate=0,align=center]{\footnotesize deplacement}; 
    \node at (-5.9cm,-1.75cm)[rotate=0,align=center]{\footnotesize extract};  
    \node at (-5.9cm,-2.05cm)[rotate=0,align=center]{\footnotesize Rx pilots};  
    \node at (-4.1cm,-3.8cm)[rotate=0,align=center]{\footnotesize $M$};
    \node at (-5.5cm,-5.3cm)[rotate=0,align=center]{\footnotesize $N$};
    \node at (0.3cm,-3.8cm)[rotate=0,align=center]{\footnotesize $M$};
    \node at (-1.1cm,-5.3cm)[rotate=0,align=center]{\footnotesize $N$};
    \node at (-0.9cm,-0.4cm)[rotate=0,align=center]{\footnotesize $M$};
    \node at (-2.4cm,-1.9cm)[rotate=0,align=center]{\footnotesize $N$};
    \node at (4.45cm,-3.8cm)[rotate=0,align=center]{\footnotesize $M'$};
    \node at (3.1cm,-4.9cm)[rotate=0,align=center]{\footnotesize $N'$};
    \node[rectangle,draw = lightgray,text = black,fill = green!3] (r) at (-2.4cm,-0.5cm) {\footnotesize \Umbruch{est. CMD coef.}};
    \node[rectangle,draw = lightgray,text = black,fill = green!3] (r) at (-1.1cm,-3.7cm) {\footnotesize \Umbruch{equal. TF frame}};
    \node[rectangle,draw = lightgray,text = black,fill = green!3] (r) at (-5.5cm,-3.7cm) {\footnotesize \Umbruch{Rx TF frame}};
    \node[rectangle,draw = lightgray,text = black,fill = green!3] (r) at (3.1cm, -3.7cm) {\scriptsize \Umbruch{precoded data frame}};
  \end{tikzpicture}}
  \caption{Structure of lineally precoded multicarrier systems including channel estimation.}
  \label{fig:overvies}  
\end{figure*} 


\section{Accordion pilot placement}
\label{section-accordion}
In this section, we introduce the proposed \emph{accordion pilot placement}.
Let us explain the choice of this name.
At the transmitter, the precoded data frame is spread out to place pilots between the data symbols. 
This is required to properly estimate the channel.
At the receiver, the pilots are then extracted and 
despreading is applied to obtain the initial data frame. 
This procedure is similar to the movement of an accordion and explains its naming.
The fundamental idea of the proposed pilot placement is to use a fixed amount of pilots $P'$ in each row (or column) and successively shift the positions circularly by some fixed hop size $\mu \in \Z$.
We have to carefully choose a suitable hop size, otherwise pilots remain clustered.
\subsection{General idea of using lattices}
\label{subsection-idea}
To explain the idea behind finding a suitable candidate for the shift $\mu$, we assume for now that $N'$ is divisible by $P'$.
Then, we can construct $\mathcal P$ from a lattice on $\Z^2$
of the form
\begin{equation}
    \Lambda^{\lambda, \mu} = \left\{ \left[ \ell, \lambda k + \mu \ell \right]^\intercal : k, \ell \in \Z \right\}
\end{equation}
and consider the restriction of the pilot indices by
\begin{equation}
    \mathcal P = \Lambda^{\lambda, \mu} \cap \mathcal I,
\end{equation}
where $\lambda = (N' + P') / P'$ is the distance between two pilots within each row and $\mu \in \Z$ is the circular shift from row to row.
We target to find the most appropriate $\mu$. 
For a given set $\mathcal P$, we consider the minimal distance between mutually distinct points given by
\begin{equation}
    d(\mathcal P) 
    = \min_{u \neq v \in \mathcal P} \|u - v\|_2.
\end{equation}
We may say $\mathcal P$ is uniformly distributed in the index grid $\mathcal I$, if it maximizes the minimal distance, i.e., $\mathcal P$ is a solution to 
\begin{equation}
    \max_{\mathcal P \subset \mathcal I} \ d(\mathcal P), \quad \text{s.t.} \quad \# \mathcal P = P.
    \label{eq-optimal-pilot-placement}
\end{equation}

Unfortunately, $d(\mathcal P)$ can not be computed easily, but $d(\Lambda^{\lambda, \mu})$ can.
We therefore rather solve
\begin{equation}
    \max_{\mu \in \Z} \ d(\Lambda^{\lambda,\mu}).
\end{equation}
Note that $\Lambda^{\lambda, \mu}$ contains $0$ and we simply have
\begin{align}
    d(\Lambda^{\lambda,\mu})^2
    &= \min_{0 \neq v \in \Lambda^{\lambda, \mu}} \|v\|_2^2
    \label{eq-min-dist-1}
    \\ &= \min_{0 \neq (k, \ell) \in \Z^2} \ell^2 + (\lambda k + \mu \ell)^2.
    \label{eq-min-dist-2}
\end{align}
Computing the squared minimal distance is actually a quadratic integer optimization problem.
For fixed $\ell \in \Z$, it is easy to compute a minimizer $k_{\mu, \ell}$ like 
\begin{equation}
    k_{\mu, \ell} = \begin{cases}
        \operatorname{round}({\mu \ell} / {\lambda}), & \text{if } \ell \neq 0, \\
        \pm 1, & \text{if } \ell = 0.
        \end{cases}
        \label{eq-kell}
\end{equation}
Moreover, any solution $(k, \ell)$ to \cref{eq-min-dist-2} satisfies
\begin{equation}
    \ell^2 \leq \ell^2 + (\lambda k  + \mu \ell)^2
    \leq d(\Lambda^{\lambda, \mu})^2 \leq \lambda^2,
\end{equation}
i.e., we have $\ell \in \{-\lambda, \dots, \lambda\}$.
It therefore suffices to compute
\begin{equation}
    d(\Lambda^{\lambda, \mu})^2 
    = \min_{\ell = -\lambda, \dots, \lambda} \ell^2 + (\lambda k_{\mu, \ell}  + \mu \ell)^2,
\end{equation}
which is fast to compute for any given $\mu$.
Finally, because $\Lambda^{\lambda, \mu} = \Lambda^{\lambda, \mu + \lambda}$ for all $\lambda \in \Z$, we can restrict the search space to $\mu = 0, \dots, \lambda -1$ and obtain the optimal shift by solving
\begin{equation}
    \mu_\text{opt} = \argmax_{\mu = 0, \dots, \lambda-1} \ \min_{\ell = -\lambda, \dots, \lambda} \ell^2 + (\lambda k_{\mu, \ell} + \mu \ell)^2.
    \label{eq-beta-opt}
\end{equation}

\subsection{General algorithm}
\label{subsection-accodion-algorithm}
\begin{figure}[t]
    \centering
    \includegraphics[width=0.23\textwidth]{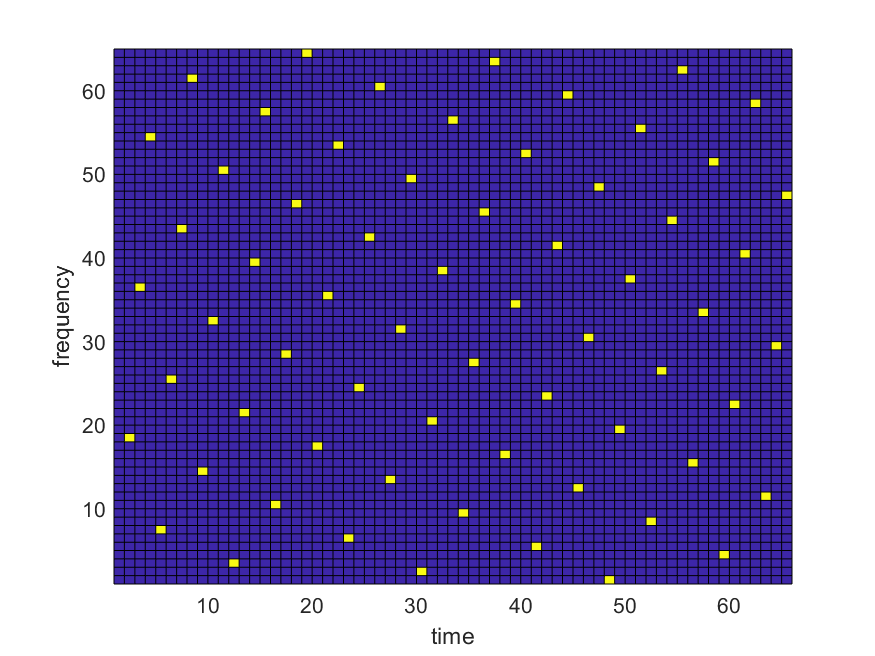}
    \centering
    \includegraphics[width=0.23\textwidth]{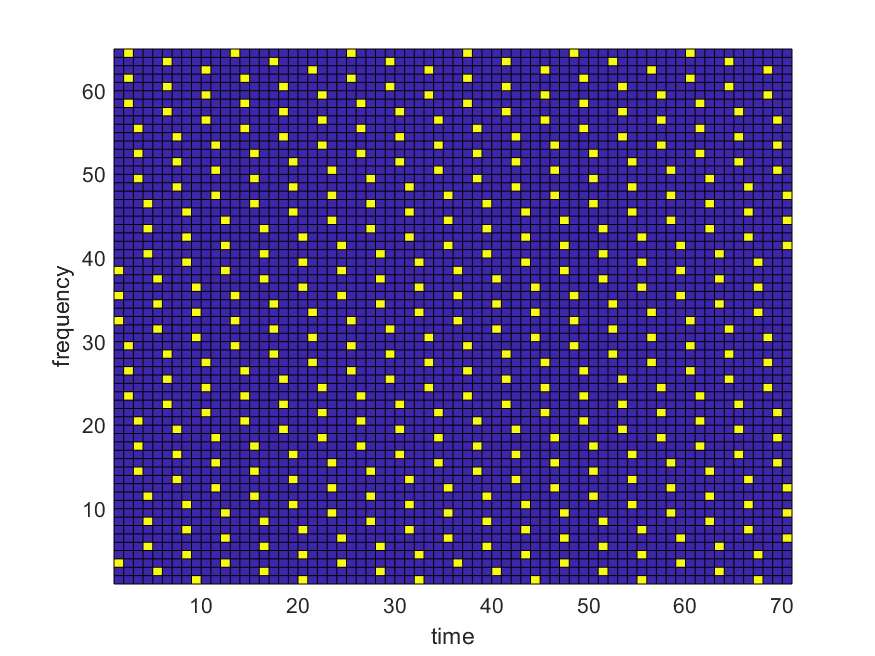}
    \caption{Generated pilot placement according to \Cref{algo-accordion} for $M' = N' = 64$ and $P'= 1$ (left) and $P' = 6$ (right). The transmit frame size is $64 \times 65$ (left) and $64 \times 70$ (right).}
    \label{figure-accordion}
\end{figure}

We have seen how to construct $\mathcal P$ in an ideal case, i.e., if the $P'$ pilots distribute uniformly along each row of length $N = N' + P'$.
The general algorithm to find a fitting accordion placement is given in
\Cref{algo-accordion}.
The key idea is to determine the ideal shift value for an approximate lattice by rounding $\lambda$ first and then computing $\mu $ according to \cref{eq-beta-opt} for the idealized setting, as seen in \cref{step-N,step-M,step-alpha,step-beta} of \Cref{algo-accordion}.
In \cref{step-row}, we place pilots as uniformly as possible in a single row using rounding.
From \cref{step-loop} onwards, we take the row indices $\mathcal R$ and shift them circularly by $\mu$ and append the new indices to $\mathcal P$.
An example of the proposed accordion placement is shown in Fig.~\ref{figure-accordion}.

\begin{algorithm}[b]
    \caption{Accordion pilot placement\label{algo-accordion}}
    \begin{algorithmic}[1]
        \Require Data frame size $(M', N')$ and pilots per row $P'$.
        \Ensure Pilot indices $\mathcal P \subset \mathcal I$, transmit frame size $(M, N)$.
        \State $M \gets M'$ \label{step-N}
        \State $N \gets N' + P'$ \label{step-M}
        \State $\lambda \gets \operatorname{round}(N / P')$ \label{step-alpha}
        \State $\displaystyle \mu_\text{opt} \gets \argmax_{ \mu = 0, \dots, \lambda-1} \ \min_{\ell = -\lambda, \dots, \lambda} \ell^2 + (\lambda k_{ \mu, \ell} +  \mu \ell)^2$
        \label{step-beta}
        \vspace{0.2em}
        \Statex \hspace{1.8em} where $k_{ \mu, \ell}$ as defined in \cref{eq-kell}.
        \vspace{0.25em}
        \State $\mathcal R \gets \left\{ \operatorname{round}(\bar n N / P') : \bar n = 0, \dots, P'-1 \right\}$
        \label{step-row}
        \State $\mathcal P^{(0)} \gets \varnothing$
        \For{$\bar m = 1, \dots, M$} \label{step-loop}
            \State $\mathcal R^{(\bar m)} \gets \left(\mu \bar m + \mathcal R\right) \mod{N}$
            \State $\mathcal P^{(\bar m)} \gets \mathcal P^{(\bar m -1)} \cup \left\{ (\bar m, \bar n) : \bar n \in \mathcal R^{(\bar m)} \right\}$
        \EndFor
        \State $\mathcal P \gets \mathcal P^{(M)}$
    \end{algorithmic}
\end{algorithm}

\section{Numerical simulations and results}
\label{section-numerics}
In this section, we present the simulation setup and numerical results. 
We compare the performance of different linear precoding transformations and
evaluate the proposed channel estimation scheme. 
\subsection{Numerical simulation setup}
\label{subsection-numeric-setup}
For the numerical evaluation, we choose a typical \gls{urllc} scenario in which a vehicle receives short-frame messages from a base station \cite{PfadlerWSA2020}.
In this scenario, the vehicle has to reliably recover the transmitted bits from each frame. To do so, the channel is estimated and equalized on a per frame basis.
\par
To evaluate pulse-shaped multicarrier systems in high-mobility scenarios, the right choice of the simulation setup is essential. 
In Table~\ref{simulationparam}, we list the selected simulation and system parameters.
We use the geometric-statistical channel simulator QuaDRiGa to generate the channels  \cite{jaeckel2014quadriga}. 
To obtain doubly-dispersive channels, we update the channel samples generated by the simulator within the duration of one frame. 
The channel then becomes time-variant at higher velocities.
We therefore configure the simulator to update the channel samples at a rate of $\nicefrac{1}{B}$=0.2\,$\mu$s. 
For the  channel model, we choose the 3GPP 38.901 UMi \gls{nlos} model which takes $R$=58 multipaths into account.
All together, the high sampling rate, the \gls{nlos} scenario, and high velocities allow us to obtain highly time-variant channels with our simulation setup.  
\par
To realize the pulse-shaped multicarrier system, we use \mbox{LTFAT} which provides a transceiver structure based on a polyphase implementation of filtering \cite{ltfatnote030}. 
We choose a \gls{tf} product of \mbox{$TF=1.25$} to balance the trade-off between the signal to interference ratio and spectral efficiency \cite{matz2007analysis}.
In the \gls{tf} domain, we design the short-frame to consist of $N=64$ time steps and $M=64$ frequency steps \cite{PfadlerWCL}.
We use a bandwidth of $B=5$\,MHz and a frame duration of $T_f=1$\,ms.
This results in a time step size of $T=16\,\mu$s and a frequency step size of $F=78.125$\,kHz which are also referred to as symbol length and subcarrier spacing, respectively. 
At the Gabor filterbank, we utilize orthogonalized Gaussian-like pulses to synthetize and analyze the transmitted and received signal in the time domain, respectively. 
These pulses are generated by orthogonalizing a prototype pulse on a tight Gabor frame which is commonly referred to as \mbox{$S^{-1/2}$-trick} \cite{jung2007wssus, Sahin2014}.\footnote{canonical tight Gabor frame \url{http://ltfat.org/doc/gabor/gabtight.html}}
These pulses are identical by construction, i.e., $\gamma=g$, and each pulse is orthogonal to its translations on the \gls{tf} grid.
From the channel simulator, we get 58 multipaths with 5120 time samples for each of them.
The total length of these channel samples corresponds to the duration of one frame, i.e., $T_f$=1\,ms.
Then, we apply a time-varying convolution between the transmit signal and each multipath and obtain the superposition of all of them as received signal.
To assure a cyclic convolution, we add a block cyclic prefix to the samples with appropriate length. 
%
\begin{table}[b]
\caption{Simulation and system parameters\label{simulationparam}}
\centering
\scriptsize
\begin{tabular}{r|c|l}
Parameter									&Notation										&Value/description 											\\
\hline\hline
Carrier frequency							& $f_{c}$										&	5.9\,GHz						\\ \hline
Bandwidth									& $B$												& 5\,MHz							\\ \hline
Frame duration								& $T_f$												& 1\,ms                     \\ \hline
Time step size			& $T$												& 16\,$\mu$s									\\ \hline
Frequency step size				& $F$												& 78.125\,kHz									\\ \hline
Number of time steps			& $N$												& 64									\\ \hline
Number of frequency steps				& $M$												&64									\\ \hline
Modulation scheme							& -											& QPSK										\\ \hline
Time-frequency product 						& $TF$											& 1.25								\\ \hline
Synthesis and analysis pulse&			$\gamma$ = $g$				& orthogonalized Gaussian-like										\\ \hline
Channel simulator 					&-						& QuaDRIGa v2.4.0 \cite{jaeckel2014quadriga}										\\ \hline
Channel model 					&	-					& 3GPP 38.901 UMi \gls{nlos}									\\ \hline
\end{tabular}
\end{table}

\begin{figure}[t]
\captionsetup[subfloat]{farskip=6pt,captionskip=1pt}
\centering
\subfloat[Relative symbol mean square error]{
    \begin{tikzpicture}
    \footnotesize
	    \begin{axis}[
    	    xmax=400,%
    	    xmin=12.5,%
    	    xtick={25,50,100,200,300,400},
    	    height = 5cm,
    	    width=0.49\textwidth,
    	    xlabel=relative velocity ($\nicefrac{\text{km}}{\text{h}}$),
    	    ylabel=MSE (dB),
    	    grid=both,
    	    mark repeat = 1,
    	    legend style={at={(0.65,0.02)},anchor=south west}, mark options={solid, draw = black}, mark size = 1.5pt]
    	    \addplot+ [mark=*,line width=1pt]  table [y=none, x=v]{figs/sym_mse_file_12db.dat};
    	    \addlegendentry{\tiny none}
    	    \addplot+ [mark = triangle,line width=1pt]  table [y=dsftsub2, x=v]{figs/sym_mse_file_12db.dat};
    	    \addlegendentry{\tiny \gls{dsft}-SF-2}
    	    \addplot+ [mark = x,line width=1pt]  table [y=dsftsub4, x=v]{figs/sym_mse_file_12db.dat};
    	    \addlegendentry{\tiny \gls{dsft}-SF-4}
    	    \addplot+ [black,mark=diamond*,line width=1pt]  table [y=dsftsub8, x=v]{figs/sym_mse_file_12db.dat};
    	    \addlegendentry{\tiny \gls{dsft}-SF-8}
    	    \addplot+ [mark = triangle,line width=1pt]  table [y=fft, x=v]{figs/sym_mse_file_12db.dat};
    	    \addlegendentry{\tiny 1D FFT}
    	    \addplot+ [mark = x,line width=1pt]  table [y=fft2, x=v]{figs/sym_mse_file_12db.dat};
    	    \addlegendentry{\tiny 2D FFT}
    	    \addplot+ [mark = diamond*,line width=1pt]  table [y=dsft, x=v]{figs/sym_mse_file_12db.dat};
    	    \addlegendentry{\tiny 1D \gls{dsft}}
    	    \addplot+ [mark = |,line width=1pt]  table [y=fwht, x=v]{figs/sym_mse_file_12db.dat};
    	    \addlegendentry{\tiny 1D \gls{fwht}}
    	    \addplot+ [mark = otimes*,line width=1pt]  table [y=dsft, x=v]{figs/sym_mse_file_12db.dat};
    	    \addlegendentry{\tiny 2D \gls{dsft}}
    	    \addplot+ [ mark = triangle,line width=1pt]  table [y=fwht2, x=v]{figs/sym_mse_file_12db.dat};
    	    \addlegendentry{\tiny 2D \gls{fwht}}
    	    \addplot+ [ mark = triangle,line width=1pt]  table [y=fwht2, x=v]{figs/sym_mse_file_12db.dat};
    	    \addlegendentry{\tiny random}
	        \node at (axis cs: 140,-9.58) (pt2) {};
	        \node at (axis cs: 70, -9.5) (pt1) {};
	        \node (pt2edge) at ($(pt2)+(110:0.15cm)$)  {};
	        \draw (pt1.center) -- (pt2edge.center);
	        \node [font = \scriptsize, yshift=3pt] at (pt1.center) {\text{all}};
	        \draw[index of colormap=5,thick] (pt2.center) circle [radius=0.15cm];
	    \end{axis}
    \end{tikzpicture}
\label{fig:op:mse}}
\hfill
\subfloat[Uncoded bit error rate]{
    \begin{tikzpicture}
    \footnotesize
	    \begin{axis}[
    	    xmax=400,%
    	    xmin=12.5,%
    	    xtick={25,50,100,200,300,400},
    	    ytick={-17,-18,-19,-20,-21,-22,-23,-24,-25},
    	    height = 5cm,
    	    width=0.49\textwidth,
    	    xlabel=relative velocity ($\nicefrac{\text{km}}{\text{h}}$),
    	    ylabel=BER (dB),
    	    grid=both,
    	    mark repeat = 1,
    	    legend style={at={(0.65,0.3)},anchor=south west}, mark options={solid, draw = black}, mark size = 1.5pt]
    	    mark repeat = 1,
    	    legend style={at={(0.57,0.24)},anchor=south west}, mark options={solid, draw = black}, mark size = 1.5pt]
    	    \addplot+ [mark=*,line width=1pt]  table [y=none, x=v]{figs/ber_file_12db.dat};
    	    \addplot+ [mark = triangle,line width=1pt]  table [y=dsftsub2, x=v]{figs/ber_file_12db.dat};
    	    \addplot+ [mark = x,line width=1pt]  table [y=dsftsub4, x=v]{figs/ber_file_12db.dat};
    	    \addplot+ [black,mark=diamond*,line width=1pt]  table [y=dsftsub8, x=v]{figs/ber_file_12db.dat};
    	    \addplot+ [mark = triangle,line width=1pt]  table [y=fft, x=v]{figs/ber_file_12db.dat};
    	    \addplot+ [mark = x,line width=1pt]  table [y=fft2, x=v]{figs/ber_file_12db.dat};
    	    \addplot+ [mark = diamond*,line width=1pt]  table [y=dsft, x=v]{figs/ber_file_12db.dat};
    	    \addplot+ [mark = |,line width=1pt]  table [y=fwht, x=v]{figs/ber_file_12db.dat};
    	    \addplot+ [mark = otimes*,line width=1pt]  table [y=dsft, x=v]{figs/ber_file_12db.dat};
    	    \addplot+ [ mark = triangle,line width=1pt]  table [y=fwht2, x=v]{figs/ber_file_12db.dat};
	       	\node at (axis cs: 140, -32.5+8.3) (pt2) {};
	        \node at (axis cs: 200, -31+7.5) (pt1) {};
	        \node (pt2edge) at ($(pt2)+(110:0.15cm)$)  {};
	        \draw (pt1.center) -- (pt2edge.center);
	        \node [font = \scriptsize, yshift=3pt] at (pt1.center) {\text{all remaining}};
	        \draw[index of colormap=5,thick] (pt2.center) circle [radius=0.15cm];
	    \end{axis}
    \end{tikzpicture}
\label{fig:op:ber}}
\hfill
\subfloat[Normalized maximal symbol-error deviation]{
    \begin{tikzpicture}
    \footnotesize
	    \begin{axis}[
    	    xmax=400,%
    	    xmin=12.5,%
    	    xtick={25,50,100,200,300,400},
    	    height = 5cm,
    	    width=0.49\textwidth,
    	    xlabel=relative velocity ($\nicefrac{\text{km}}{\text{h}}$),
    	    ylabel=NMSED (dB),
    	    grid=both,
    	    mark repeat = 1,
    	    legend style={at={(0.57,0.24)},anchor=south west}, mark options={solid, draw = black}, mark size = 1.5pt]
    	    \addplot+ [mark=*,line width=1pt]  table [y=none, x=v]{figs/sym_papr_file_12db-v1.dat};
    	    \addplot+ [mark = triangle,line width=1pt]  table [y=dsftsub2, x=v]{figs/sym_papr_file_12db-v1.dat};
    	    \addplot+ [mark = x,line width=1pt]  table [y=dsftsub4, x=v]{figs/sym_papr_file_12db-v1.dat};
    	    \addplot+ [black,mark=diamond*,line width=1pt]  table [y=dsftsub8, x=v]{figs/sym_papr_file_12db-v1.dat};
    	    \addplot+ [mark = triangle,line width=1pt]  table [y=fft, x=v]{figs/sym_papr_file_12db-v1.dat};
    	    \addplot+ [mark = x,line width=1pt]  table [y=fft2, x=v]{figs/sym_papr_file_12db-v1.dat};
    	    \addplot+ [mark = diamond*,line width=1pt]  table [y=dsft, x=v]{figs/sym_papr_file_12db-v1.dat};
    	    \addplot+ [mark = |,line width=1pt]  table [y=fwht, x=v]{figs/sym_papr_file_12db-v1.dat};
    	    \addplot+ [mark = otimes*,line width=1pt]  table [y=dsft, x=v]{figs/sym_papr_file_12db-v1.dat};
    	    \addplot+ [ mark = triangle,line width=1pt]  table [y=fwht2, x=v]{figs/sym_papr_file_12db-v1.dat};
	        \node at (axis cs: 150,9.25) (pt2) {};
	        \node at (axis cs: 200, 9.4) (pt1) {};
	        \node (pt2edge) at ($(pt2)+(110:0.15cm)$)  {};
	        \draw (pt1.center) -- (pt2edge.center);
	        \node [font = \scriptsize, yshift=3pt] at (pt1.center) {\text{all remaining}};
	        \draw[index of colormap=5,thick] (pt2.center) circle [radius=0.15cm];
	    \end{axis}
    \end{tikzpicture}
\label{fig:op:papr:nlos}}
\hfill
\caption{Performance comparison of different linear precoding transformations at 12\,dB \gls{snr} assuming full \gls{csi} of \gls{cmd}.}
\label{fig:performance:op}  
\end{figure}
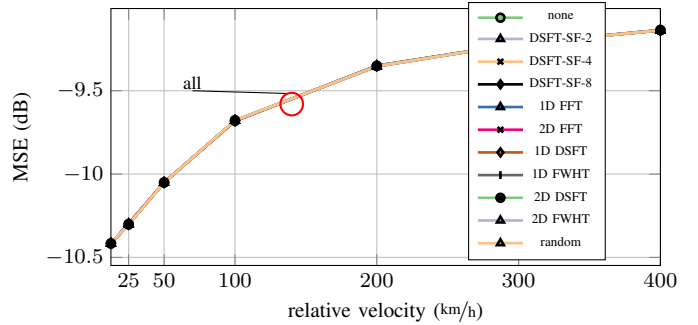
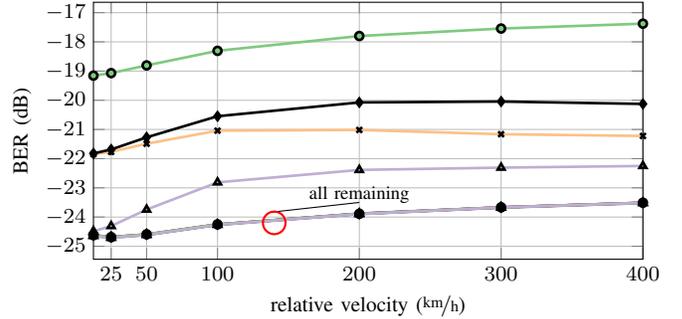
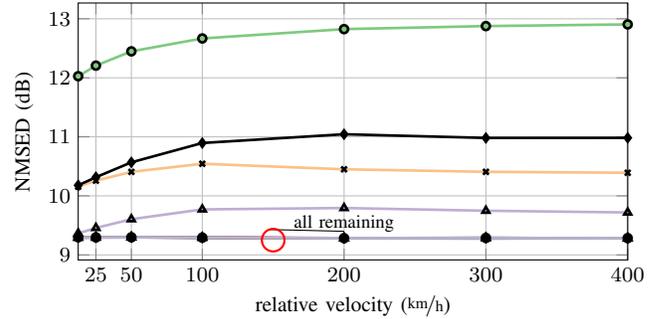
\begin{figure*}[!t]
\captionsetup[subfloat]{farskip=1pt,captionskip=1pt}
\centering
\subfloat[Uncoded \gls{ber} using 256 pilot symbols]{
      \centering
    \begin{tikzpicture}
    \footnotesize
	    \begin{axis}[
    	    xmax=24,%
    	    xmin=0,%
    	    ymax=0,%
    	    ymin=-50,%
    	    xtick={0,2,4,6,8,10,12,14,16,18,20,22,24,26},
    	    ytick={0,-5,-10,-15,-20,-25,-30,-35,-40,-45,-50},
    	    width=0.49\textwidth,
            height=5.5cm,
    	    xlabel=SNR (dB),
    	    ylabel=BER (dB),
    	    grid=both,
    	    mark repeat = 2,
    	    legend style={at={(0.01,0.01)},anchor=south west}, mark options={solid, draw = black}, mark size = 1.5pt]
    	    \addplot+ [mark phase = 1, index of colormap=1 of Accent-8, mark = star,line width=1.5pt]  table [y=LQ, x=SNR]{uncoded_ber_npilots_256_speed_200.dat};
    	    \addlegendentry{\scriptsize LMMSE}
    	    \addplot+ [mark phase = 1, index of colormap=4 of Accent-8, mark = |,line width=1.5pt]  table [y=D2, x=SNR]{uncoded_ber_npilots_256_speed_200.dat};
    	    \addlegendentry{\scriptsize SRH}
    	    \addplot+ [mark phase = 2, index of colormap=8 of Accent-8, mark = triangle,line width=1.5pt]  table [y=D2snr, x=SNR]{uncoded_ber_npilots_256_speed_200.dat};
    	    \addlegendentry{\scriptsize SRH-NA}
    	    \addplot+ [mark phase = 0, index of colormap=5 of Accent-8, mark = x,line width=1.5pt]  table [y=D2aniso, x=SNR]{uncoded_ber_npilots_256_speed_200.dat};
    	    \addlegendentry{\scriptsize SRH-MA}
    	    \addplot+ [mark phase = 4, index of colormap=6 of Accent-8, mark = diamond*,line width=1.5pt]  table [y=D2anisosnr, x=SNR]{uncoded_ber_npilots_256_speed_200.dat};
    	    \addlegendentry{\scriptsize SRH-MNA}
    	    \addplot+ [mark phase = 3, index of colormap=7 of Accent-8, mark = o,line width=1.5pt]  table [y=ideal, x=SNR]{uncoded_ber_npilots_256_speed_200.dat};
    	    \addlegendentry{\scriptsize perfect \gls{cmd}}
    	    %
    	    \coordinate (A) at (11.5,-20);
            \coordinate (B) at (17,-20);
            \draw[line width=0.3mm,|-|]  (A) --  (B); 
            \draw[] (A) to node[above] {\scriptsize $5.5$\,dB} (B);
	    \end{axis}
    \end{tikzpicture}
    \label{fig-nlos-uncoded-256}}
\hfill
\subfloat[Coded \gls{ber} (convolutional hard-decision decoding at rate of \mbox{$\nicefrac{1}{3}$ )} using 
256 pilot symbols.]{
      \centering
    \begin{tikzpicture}
    \footnotesize
	    \begin{axis}[
    	    xmax=24,%
    	    xmin=0,%
    	    ymax=0,%
    	    ymin=-50,%
    	    xtick={0,2,4,6,8,10,12,14,16,18,20,22,24,26},
    	    ytick={0,-5,-10,-15,-20,-25,-30,-35,-40,-45,-50},
    	    width=0.49\textwidth,
            height=5.5cm,
    	    xlabel=SNR (dB),
    	    ylabel=BER (dB),
    	    grid=both,
    	    mark repeat = 2,
    	    legend style={at={(0.57,0.65)},anchor=south west}, mark options={solid, draw = black}, mark size = 1.5pt]
    	    \addplot+ [mark phase = 1, index of colormap=1 of Accent-8, mark = star,line width=1.5pt]  table [y=LQ, x=SNR]{coded_ber_npilots_256_speed_200.dat};
    	    \addplot+ [mark phase = 1, index of colormap=4 of Accent-8, mark = |,line width=1.5pt]  table [y=D2, x=SNR]{coded_ber_npilots_256_speed_200.dat};
    	    \addplot+ [mark phase = 2, index of colormap=8 of Accent-8, mark = triangle,line width=1.5pt]  table [y=D2snr, x=SNR]{coded_ber_npilots_256_speed_200.dat};
    	    \addplot+ [mark phase = 0, index of colormap=5 of Accent-8, mark = x,line width=1.5pt]  table [y=D2aniso, x=SNR]{coded_ber_npilots_256_speed_200.dat};
    	    \addplot+ [mark phase = 4, index of colormap=6 of Accent-8, mark = diamond*,line width=1.5pt]  table [y=D2anisosnr, x=SNR]{coded_ber_npilots_256_speed_200.dat};
    	    \addplot+ [mark phase = 3, index of colormap=7 of Accent-8, mark = o,line width=1.5pt]  table [y=ideal, x=SNR]{coded_ber_npilots_256_speed_200.dat};
	    \end{axis}
    \end{tikzpicture}
\label{fig-nlos-coded-256}}
\hfill
\subfloat[Uncoded \gls{ber} using 512 pilot symbols]{
    \centering
    \begin{tikzpicture}
    \footnotesize
	    \begin{axis}[
    	    xmax=24,%
    	    xmin=0,%
    	    ymax=0,%
    	    ymin=-50,%
    	    xtick={0,2,4,6,8,10,12,14,16,18,20,22,24,26},
    	    ytick={0,-5,-10,-15,-20,-25,-30,-35,-40,-45,-50},
    	    width=0.49\textwidth,
            height=5.5cm,
    	    xlabel=SNR (dB),
    	    ylabel=BER (dB),
    	    grid=both,
    	    mark repeat = 2,
    	    legend style={at={(0.57,0.65)},anchor=south west}, mark options={solid, draw = black}, mark size = 1.5pt]
    	    \addplot+ [mark phase = 1, index of colormap=1 of Accent-8, mark = star,line width=1.5pt]  table [y=LQ, x=SNR]{uncoded_ber_npilots_512_speed_200.dat};
    	    \addplot+ [mark phase = 1, index of colormap=4 of Accent-8, mark = |,line width=1.5pt]  table [y=D2, x=SNR]{uncoded_ber_npilots_512_speed_200.dat};
    	    \addplot+ [mark phase = 2, index of colormap=8 of Accent-8, mark = triangle,line width=1.5pt]  table [y=D2snr, x=SNR]{uncoded_ber_npilots_512_speed_200.dat};
    	    \addplot+ [mark phase = 0, index of colormap=5 of Accent-8, mark = x,line width=1.5pt]  table [y=D2aniso, x=SNR]{uncoded_ber_npilots_512_speed_200.dat};
    	    \addplot+ [mark phase = 4, index of colormap=6 of Accent-8, mark = diamond*,line width=1.5pt]  table [y=D2anisosnr, x=SNR]{uncoded_ber_npilots_512_speed_200.dat};
    	    \addplot+ [mark phase = 3, index of colormap=7 of Accent-8, mark = o,line width=1.5pt]  table [y=ideal, x=SNR]{uncoded_ber_npilots_512_speed_200.dat};
    	    %
    	    \coordinate (A) at (11,-20);
            \coordinate (B) at (14.7,-20);
            \draw[line width=0.3mm,|-|]  (A) --  (B); 
            \draw[] (A) to node[above] {\scriptsize $3.7$\,dB} (B);
	    \end{axis}
    \end{tikzpicture}
    \label{fig-nlos-uncoded-512}}
\hfill
\subfloat[Coded \gls{ber} (convolutional hard-decision decoding at rate of \mbox{$\nicefrac{1}{3}$ )} using 
512 pilot symbols.]{
      \centering
    \begin{tikzpicture}
    \footnotesize
	    \begin{axis}[
    	    xmax=24,%
    	    xmin=0,%
    	    ymax=0,%
    	    ymin=-50,%
    	    xtick={0,2,4,6,8,10,12,14,16,18,20,22,24,26},
    	    ytick={0,-5,-10,-15,-20,-25,-30,-35,-40,-45,-50},
    	    width=0.49\textwidth,
            height=5.5cm,
    	    xlabel=SNR (dB),
    	    ylabel=BER (dB),
    	    grid=both,
    	    mark repeat = 2,
    	    legend style={at={(0.57,0.65)},anchor=south west}, mark options={solid, draw = black}, mark size = 1.5pt]
    	    \addplot+ [mark phase = 1, index of colormap=1 of Accent-8, mark = star,line width=1.5pt]  table [y=LQ, x=SNR]{coded_ber_npilots_512_speed_200.dat};
    	    \addplot+ [mark phase = 1, index of colormap=4 of Accent-8, mark = |,line width=1.5pt]  table [y=D2, x=SNR]{coded_ber_npilots_512_speed_200.dat};
    	    \addplot+ [mark phase = 2, index of colormap=8 of Accent-8, mark = triangle,line width=1.5pt]  table [y=D2snr, x=SNR]{coded_ber_npilots_512_speed_200.dat};
    	    \addplot+ [mark phase = 0, index of colormap=5 of Accent-8, mark = x,line width=1.5pt]  table [y=D2aniso, x=SNR]{coded_ber_npilots_512_speed_200.dat};
    	    \addplot+ [mark phase = 4, index of colormap=6 of Accent-8, mark = diamond*,line width=1.5pt]  table [y=D2anisosnr, x=SNR]{coded_ber_npilots_512_speed_200.dat};
    	    \addplot+ [mark phase = 3, index of colormap=7 of Accent-8, mark = o,line width=1.5pt]  table [y=ideal, x=SNR]{coded_ber_npilots_512_speed_200.dat};
	    \end{axis}
    \end{tikzpicture}
\label{fig-nlos-coded-512}}
\hfill
\caption{
\gls{ber} as a function of the \gls{snr} for different channel estimation schemes at relative velocity of \mbox{$\Delta v=200\,\nicefrac{\text{km}}{\text{h}}$}.}
\label{fig:CMDE}
\end{figure*}
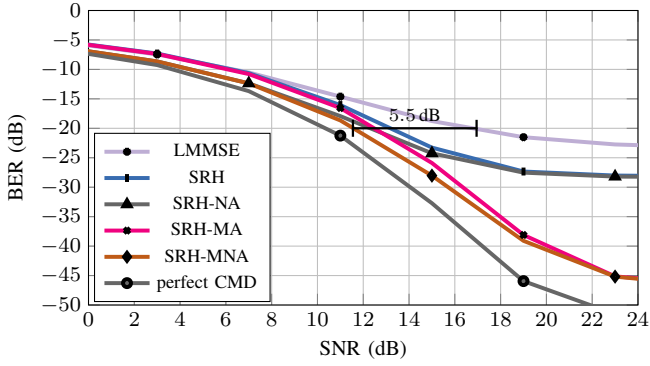
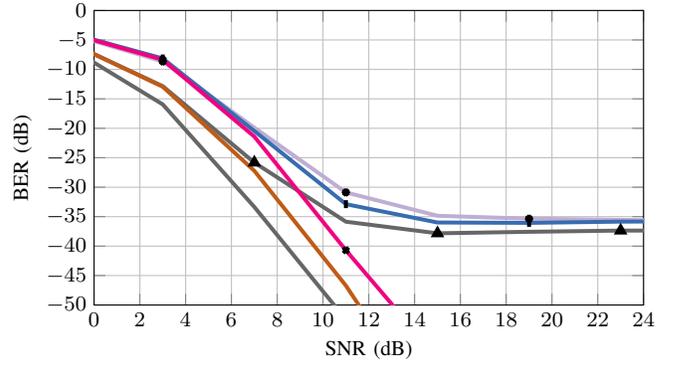
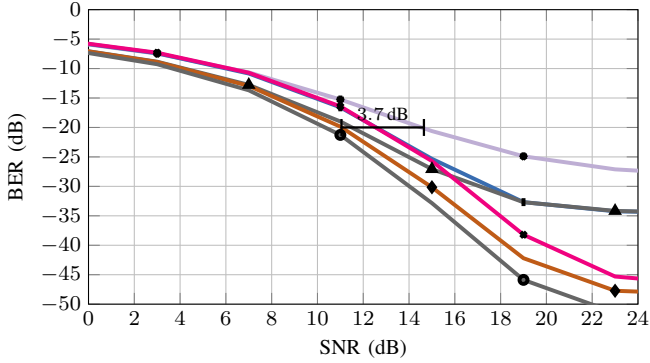
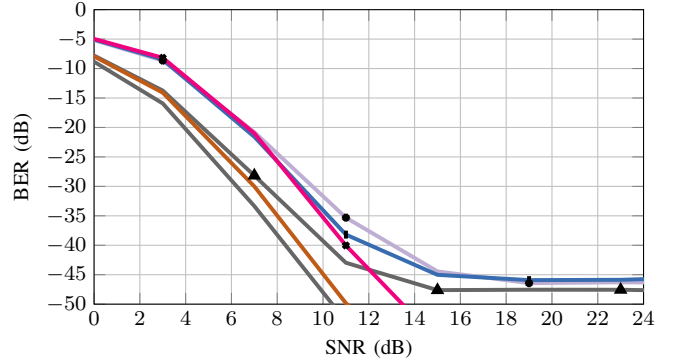
\subsection{Comparison of distinct linear precoding setups}
\label{subsection-numeric-op}
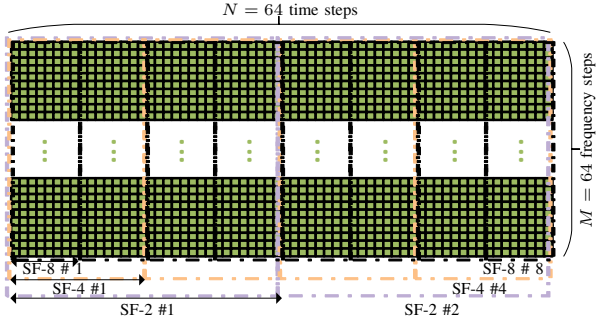
\begin{figure}[t]
  \centering 
  \scalebox{0.9}{
  \begin{tikzpicture}[]
    \node at (0cm,0cm) {\includegraphics[width=0.49\textwidth]{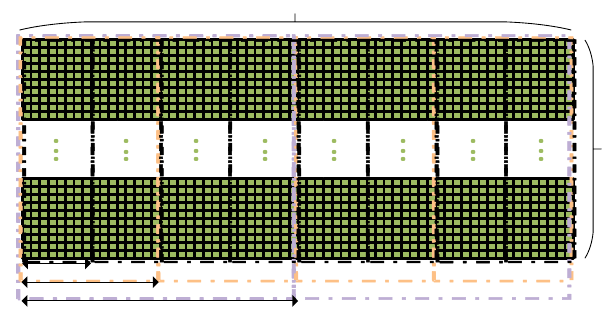}};
    \node at (4.4cm,0.3cm)[rotate=90,align=center]{\scriptsize $M=64$ frequency steps}; 
    \node at (0.0cm,2.2cm)[rotate=0,align=center]{\scriptsize $N=64$ time steps}; 
    \node at (-3.5cm,-1.6cm)[rotate=0,align=center]{\scriptsize \gls{sf}-8 \# 1};
    \node at (3.3cm,-1.6cm)[rotate=0,align=center]{\scriptsize \gls{sf}-8 \# 8};
    \node at (-3.1cm,-1.9cm)[rotate=0,align=center]{\scriptsize \gls{sf}-4 \#1};
    \node at (2.8cm,-1.9cm)[rotate=0,align=center]{\scriptsize \gls{sf}-4 \#4};
    \node at (-2.1cm,-2.2cm)[rotate=0,align=center]{\scriptsize \gls{sf}-2 \#1};
    \node at (2.1cm,-2.2cm)[rotate=0,align=center]{\scriptsize \gls{sf}-2 \#2};
  \end{tikzpicture}}
  \caption{\gls{tf} frame with three different \glsentrylong{sf} (\gls{sf}) structures.
  \gls{sf}-2 consist of two times N=32 and M=64 symbols, \gls{sf}-4 of four times N=16 and M=64 symbols, and \gls{sf}-8 of eight times N=8 and M=64 symbols. 
  In each \gls{sf} the data symbols are separately precoded.}
  \label{fig:subspace}  
\end{figure}

We numerically investigate the impact of applying different linear precoding transformations to the data frame, the gained \gls{tf} diversity, and how this gain depends on the size of precoded data frame. 
Let us therefore detail the distinct setups considered in this subsection.
Since we focus on precoding, we only place data symbols into the \gls{tf} frame and assume full \gls{csi} knowledge of the \gls{cmd} 
at the receiver.
The perfect \gls{cmd} is used for \gls{mmse} equalization as detailed in \eqref{eq:mmse}.
To precode the data frame, we apply the \gls{dsft}, \gls{fwht} and \gls{fft} each as a 1D and 2D transformation, and we consider random precoding and without precoding as well. 
In addition, we study setups in which we subdivide the \gls{tf} frame into two, four, and eight \gls{sf} corresponding to \gls{sf}-2, \gls{sf}-4, and \gls{sf}-8, respectively.
Fig.~\ref{fig:subspace} depicts the \gls{tf} frame in which the three considered \gls{sf} structures are shown.
The data frame is divided by the number of \gls{sf} into smaller data frames.
Each is then separately precoded and mapped to the corresponding \gls{sf}. 
The approach of using precoded \gls{sf} is particularly interesting for \gls{urllc}, as it provides higher flexibility.
For example, a vehicle can already process single \gls{sf} to recover the transmitted bits before receiving the entire frame.
From another perspective, the \gls{sf} can be used in multiuser scenarios improving reliability and providing higher flexibly than \gls{otfs}. 
This however comes at a price. 
Recall that by applying precoding to the data symbols, we increase the reliability of modulation or in other words we gain \gls{tf} diversity. 
This is due to the fact that equalization errors and self-interference are distributed over all symbols. 
By reducing the size of the precoded data frame, we also decrease the potential \gls{tf} diversity.
To measure the impact of \gls{tf} diversity gain on the precoded symbols, let us use the \gls{se-papr} as metric.
\par
In Fig.~\ref{fig:performance:op}, we study the \gls{mse}, the uncoded \gls{ber}, the \gls{se-papr} as a function of the velocity for all investigated setups.
Fig.~\ref{fig:op:mse} depicts the relative symbol \gls{mse} and shows that it is the same for all setups.  
Fig.~\ref{fig:op:ber} shows that all precoding functions achieve the same low \gls{ber} when applied to the entire \gls{tf} frame.
In particular, 1D and 2D precoding lead to the same \gls{ber} performance.
We observe in general significant performance gains of precoded data frames compared to data frames without precoding.
In the case of \gls{sf}, we can observe that the \gls{ber} increases by approximately 3 to 4\,dB for each subdivision of the \gls{tf} frame. 
The reason behind this behaviour can be explained when looking at \gls{se-papr} in Fig.~\ref{fig:op:papr:nlos}.  
The precoding ensures that the error energy from self-interference and equalization near zero-crossings of $\vec h$ are equally distributed across all symbols in the \gls{tf} frame.  

\begin{table}[b]
\caption{Overview of investigated channel estimator schemes\label{tabel-ES}}
\centering
\scriptsize
\begin{tabular}{r|p{6.5cm}}
Estimator					&Description\\
\hline\hline
LMMSE   					& Standard \gls{lmmse} estimator with noise-awareness\\ \hline
SRH			                & Minimizing isotropic ($\alpha=\beta$) second deviate in \eqref{eq-proposed-channel-estim} \\ \hline
SRH-NA				        & Minimizing isotropic second deviate in \eqref{eq-proposed-channel-estim} with noise-awareness by using the relaxation parameter $\delta$\\ \hline
SRH-MA			        & Minimizing the anisotropic second deviate in \eqref{eq-proposed-channel-estim} with mode-awareness by using the scaling parameter $\alpha$ and $\beta$ \\ \hline
SRH-MNA					& Combining both SRH-NA and SRH-MA\\ \hline
perfect CMD		            & Assuming full \gls{csi} of the \gls{cmd}\\ \hline
\end{tabular}
\end{table}
\subsection{Performance of channel estimation}
\label{subsection-numeric-channel-est}
Let us evaluate the performance of channel estimation in a linearly precoded multicarrier system which is shown in Fig.~\ref{fig:overvies}. 
We place both pilot and precoded data symbols into the \gls{tf} frame.
We utilize the \gls{fwht} to precode the data frame which offers reduced implementation complexity compared to \gls{dsft} precoding \cite{Bomfin2021robust}. 
The choice of precoding is however arbitrary since any other linear transformation leads to the same \gls{tf} diversity gain as shown in \Cref{subsection-numeric-op}.
The pilot symbols are placed according to \Cref{algo-accordion}. 
We estimate the \gls{cmd} and use \gls{mmse} equalization to revert the distortion incurred by doubly-dispersive channel. 
We study the proposed channel estimation scheme by solving the optimization problem in \eqref{eq-proposed-channel-estim} in four different configurations and denote them as follows: \gls{srh}, \gls{srh-na}, \gls{srh-ma}, and \gls{srh-mna}. 
This allows us to study the impact of taking noise-awareness as well as the mode-awarness into account. 
We detail the investigated estimators in Table~\ref{tabel-ES}.
\par
\begin{figure}[!t]
\captionsetup[subfloat]{farskip=6pt,captionskip=1pt}
\centering
\subfloat[Uncode \gls{ber}  at \mbox{$\Delta v=100\,\nicefrac{\text{km}}{\text{h}}$}]{
    \centering
    \begin{tikzpicture}
    \footnotesize
	    \begin{axis}[
    	    xmax=1024,%
    	    xmin=128,%
    	    ymax=-14,%
    	    ymin=-32,%
    	    xtick={128,256,512,1024},
   	        height = 4.5cm,
    	    width=0.45\textwidth,
    	    xlabel=number of pilots,
    	    ylabel=BER (dB),
    	    grid=both,
    	    mark repeat = 2,
    	    legend style={at={(0.65,0.33)},anchor=south west}, mark options={solid, draw = black}, mark size = 1pt]
    	   \addplot+ [mark phase = 1, index of colormap=1 of Accent-8, mark = star,line width=1.5pt]  table [y=LQ, x=X]{ber_speed_snr15_speed_200.dat};
    	    \addlegendentry{\scriptsize LMMSE}
    	    \addplot+ [mark phase = 1, index of colormap=4 of Accent-8, mark = |,line width=1pt]  table [y=D2, x=X]{ber_speed_snr15_speed_100.dat};
    	    \addlegendentry{\scriptsize SRH}
    	    \addplot+ [mark phase = 2, index of colormap=8 of Accent-8, mark = triangle,line width=1pt]  table [y=D2snr, x=X]{ber_speed_snr15_speed_100.dat};
    	    \addlegendentry{\scriptsize SRH-NA}
    	    \addplot+ [mark phase = 0, index of colormap=5 of Accent-8, mark = x,line width=1pt]  table [y=D2aniso, x=X]{ber_speed_snr15_speed_100.dat};
    	    \addlegendentry{\scriptsize SRH-MA}
    	    \addplot+ [mark phase = 4, index of colormap=6 of Accent-8, mark = diamond*,line width=1pt]  table [y=D2anisosnr, x=X]{ber_speed_snr15_speed_100.dat};
    	    \addlegendentry{\scriptsize SRH-MNA}
	    \end{axis}
    \end{tikzpicture}
\label{fig:performance:100}}
\hfill
\subfloat[Uncode \gls{ber} at \mbox{$\Delta v=200\,\nicefrac{\text{km}}{\text{h}}$}]{
    \centering
    \begin{tikzpicture}
    \footnotesize
	    \begin{axis}[
    	    xmax=1024,%
    	    xmin=128,%
    	    ymax=-14,%
    	    ymin=-32,%
   	        height = 4.5cm,
    	    xtick={128,256,512,1024},
    	    width=0.45\textwidth,
    	    xlabel=number of pilots,
    	    ylabel=BER (dB),
    	    grid=both,
    	    mark repeat = 2,
    	    legend style={at={(0.47,0.52)},anchor=south west}, mark options={solid, draw = black}, mark size = 1.5pt]
    	    \addplot+ [mark phase = 1, index of colormap=1 of Accent-8, mark = star,line width=1.5pt]  table [y=LQ, x=X]{ber_speed_snr15_speed_100.dat};
    	    \addplot+ [mark phase = 1, index of colormap=4 of Accent-8, mark = |,line width=1pt]  table [y=D2, x=X]{ber_speed_snr15_speed_200.dat};
    	    \addplot+ [mark phase = 2, index of colormap=8 of Accent-8, mark = triangle,line width=1pt]  table [y=D2snr, x=X]{ber_speed_snr15_speed_200.dat};
    	    \addplot+ [mark phase = 0, index of colormap=5 of Accent-8, mark = x,line width=1pt]  table [y=D2aniso, x=X]{ber_speed_snr15_speed_200.dat};
    	    \addplot+ [mark phase = 4, index of colormap=6 of Accent-8, mark = diamond*,line width=1pt]  table [y=D2anisosnr, x=X]{ber_speed_snr15_speed_200.dat};
	    \end{axis}
    \end{tikzpicture}
\label{fig:performance:200}}
\hfill
\subfloat[Uncode \gls{ber} at \mbox{$\Delta v=400\,\nicefrac{\text{km}}{\text{h}}$}]{
    \centering
    \begin{tikzpicture}
    \footnotesize
	    \begin{axis}[
    	    xmax=1024,%
    	    xmin=128,%
    	    ymax=-14,%
    	    ymin=-32,%
    	    xtick={128,256,512,1024},
    	    height = 4.5cm,
    	    width=0.45\textwidth,
    	    xlabel=number of pilots,
    	    ylabel=BER (dB),
    	    grid=both,
    	    mark repeat = 2,
    	    legend style={at={(0.47,0.25)},anchor=south west}, mark options={solid, draw = black}, mark size = 1pt]
    	    \addplot+ [mark phase = 1, index of colormap=1 of Accent-8, mark = star,line width=1.5pt]  table [y=LQ, x=X]{ber_speed_snr15_speed_400.dat};
    	    \addplot+ [mark phase = 1, index of colormap=4 of Accent-8, mark = |,line width=1pt]  table [y=D2, x=X]{ber_speed_snr15_speed_400.dat};
    	    \addplot+ [mark phase = 2, index of colormap=8 of Accent-8, mark = triangle,line width=1pt]  table [y=D2snr, x=X]{ber_speed_snr15_speed_400.dat};
    	    \addplot+ [mark phase = 0, index of colormap=5 of Accent-8, mark = x,line width=1pt]  table [y=D2aniso, x=X]{ber_speed_snr15_speed_400.dat};
    	    \addplot+ [mark phase = 4, index of colormap=6 of Accent-8, mark = diamond*,line width=1pt]  table [y=D2anisosnr, x=X]{ber_speed_snr15_speed_400.dat};
	    \end{axis}
    \end{tikzpicture}
\label{fig:performance:400}}
\hfill
\caption{\gls{ber} as a function of the number of pilots in \gls{nlos} vehicular scenarios at a \gls{snr} of 15\,dB for distinct channel estimators.}
\label{fig:performance:all}  
\end{figure}
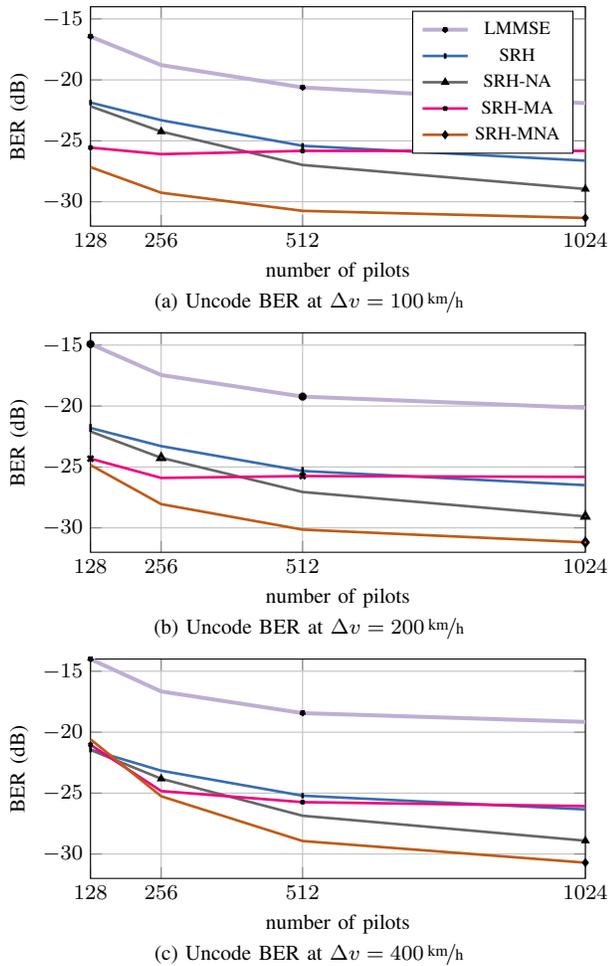

Fig.~\ref{fig:CMDE} depicts the uncoded and coded \gls{ber} as function of \gls{snr} for different numbers of pilot symbols using convolutional hard-decision decoding at rate of \mbox{$\nicefrac{1}{3}$} in the latter case.
In Fig.~\ref{fig-nlos-uncoded-256}, we show the uncoded \gls{ber} using 256 pilots. 
Fig.~\ref{fig-nlos-coded-256} illustrates that with coding an error-free transmission is only achieved for the perfect \gls{cmd}, \gls{srh-mna}, and \gls{srh-ma} estimators at a \gls{snr} of 10.5\,dB, 12\,dB, and 13\,dB, respectively.
For all remaining estimators, an error floor between $-35$ and $-38$\,dB is reached.  
Fig.~\ref{fig-nlos-uncoded-512} depicts the uncoded \gls{ber} when 512 pilot symbols are used. 
By doubling the number of pilot symbols to 512, the standard \gls{lmmse} estimator performance is improved.  
Fig.~\ref{fig-nlos-coded-512} illustrates that with coding an error-free transmission is only achieved for the perfect \gls{cmd}, \gls{srh-mna}, and \gls{srh-ma} estimator at a \gls{snr} of 10.5\,dB, 11\,dB, and 13.2\,dB, respectively.
For all remaining estimators, an error floor between $-46$ and $-48$\,dB is reached.  
Fig.~\ref{fig:performance:all} shows the uncoded \gls{ber} as a function of the number of pilot symbols at a \gls{snr} of 15\,dB with a relative velocity between 100 and 400 $\nicefrac{\text{km}}{\text{h}}$. 

 \section{Conclusions \label{sec:conclusions}}
We proposed a novel channel main diagonal estimator scheme which minimizes the energy of the second
order derivatives.
This scheme considered noise including self-interference power and the ratio of the channel spreading region for anisotropic regularization of the weighted Hessian matrix. 
The use of Tikhonov regularization allowed us to obtain an unconstrained least-squares problem where a closed form solution for each noise parameter exists and can be computed offline. 
We introduced a new pilot placement scheme that enables a more efficient use of resources and improved channel estimation.
The numerical results showed that the proposed scheme allows an accurate channel estimation and equalization of received short-frame messages even in highly time-varying communication scenarios.

\section*{Acknowledgment}
The authors acknowledge the financial support by the Federal Ministry of Education and Research of Germany in the programme of ``Souverän. Digital. Vernetzt.'' Joint project 6G-RIC, project identification number: 16KISK020K.
\bibliographystyle{ieeetr}
\bibliography{ref} 

\begin{IEEEbiography}
    [{\includegraphics[width=1in,height=1.25in,clip,keepaspectratio]{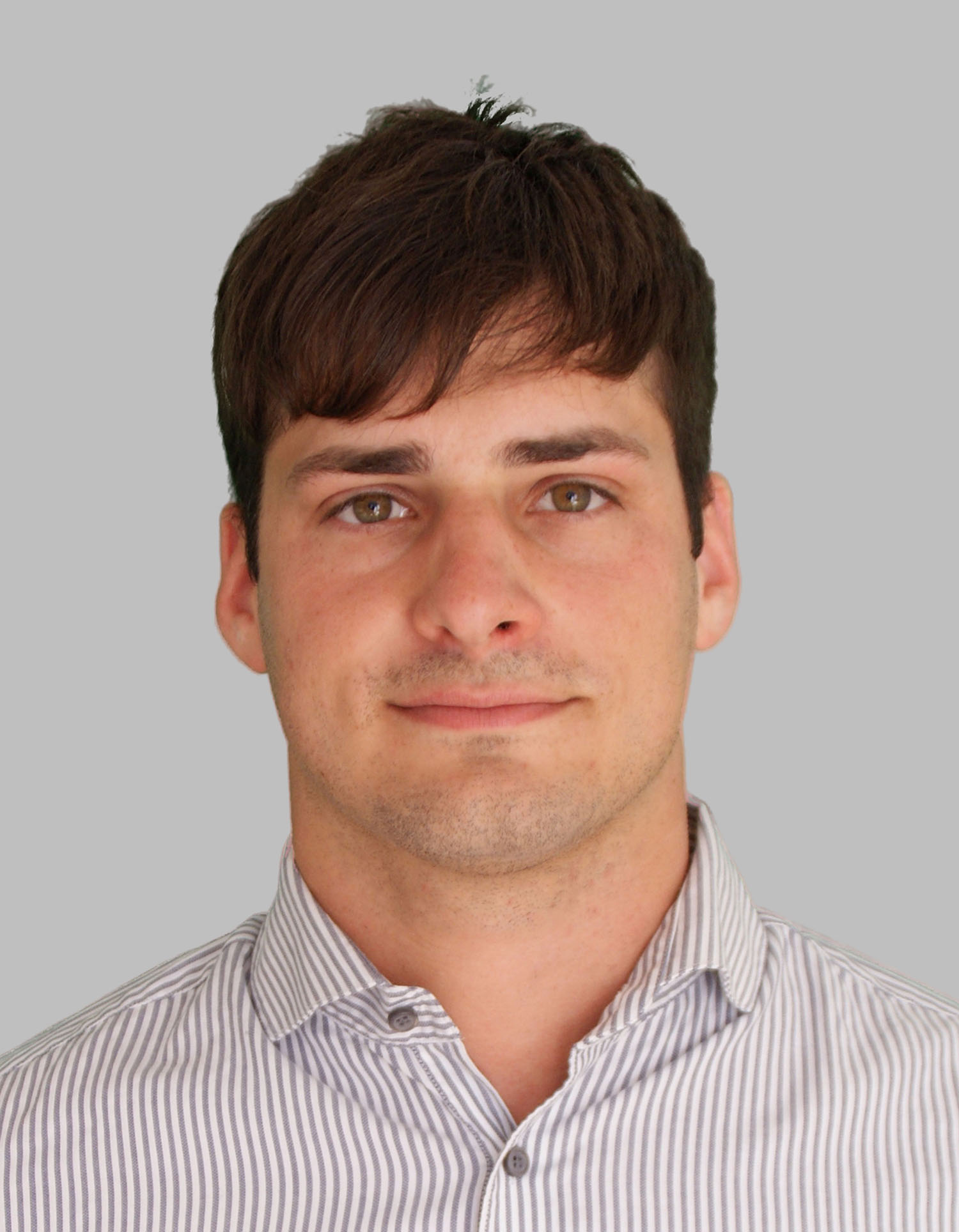}}]
    {Andreas Pfadler} received  the M.Sc. degree in Telecommunication Engineering with specialization in wireless communications from the Polytechnic University of Catalonia (UPC), Barcelona, Spain, in 2018.
He is a communication expert and function developer for automated driving at Volkswagen Commercial Vehicles and is currently pursuing the Ph.D. degree in vehicular communications with the Technische Universität of Berlin, under the supervision of Prof. Dr.-Ing. S. Stanczak.
He has been involved in several research projects as the European project 5GCroCo and the German national project 5G NetMobil. 
His research interests include antennas, signal processing, predictive quality of service, new waveforms and wave propagation.
\end{IEEEbiography}
\begin{IEEEbiography}
    [{\includegraphics[width=1in,height=1.25in,clip,keepaspectratio]{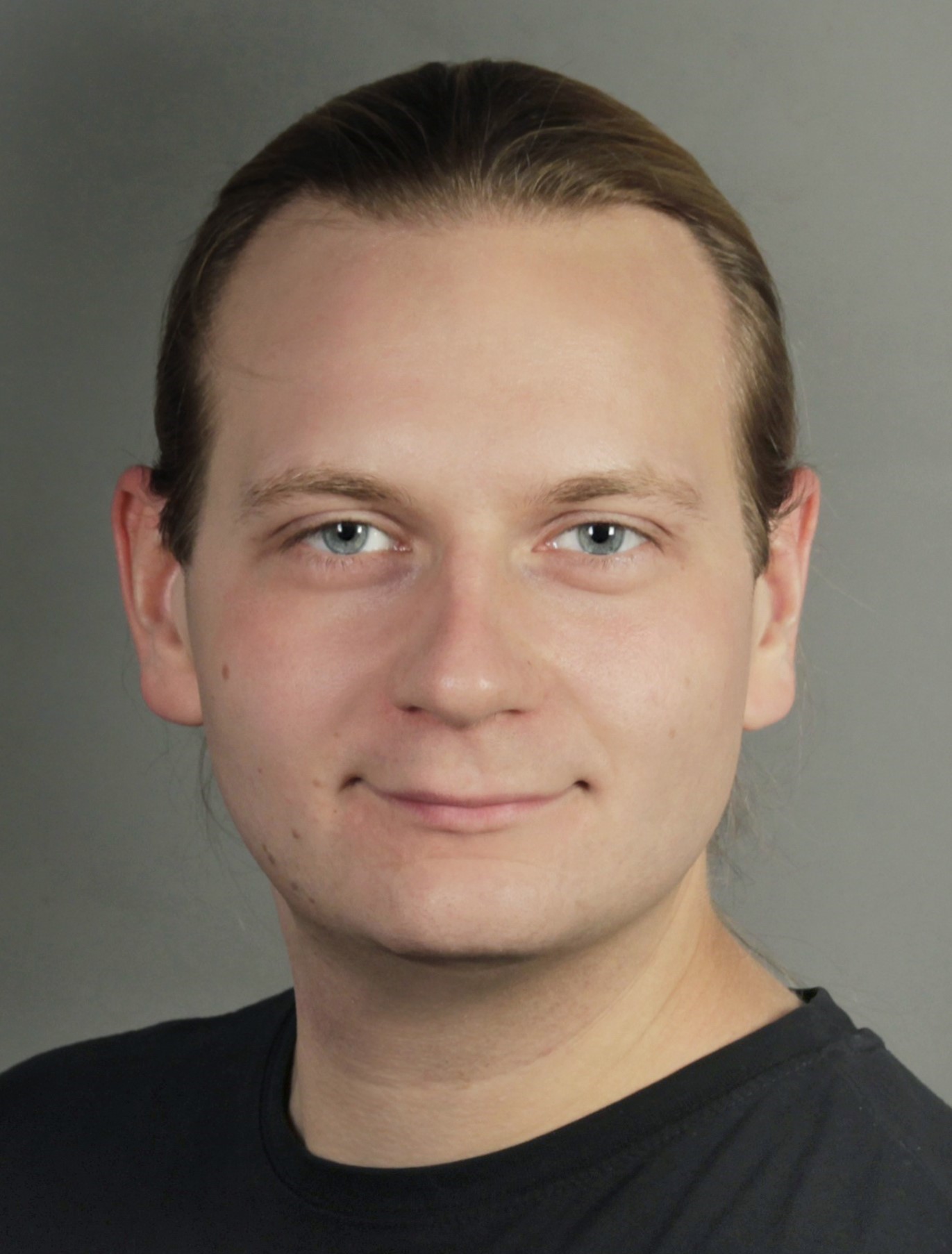}}]
    {Tom Szollmann} 
    received the M.Sc. degree in mathematics with specialization in functional analysis and probability theory from the Technical University of Berlin in 2021. He is currently pursuing the Ph.D. degree in data management of vehicular communication in the context of automated driving at Volkswagen Commercial Vehicles under the supervision of Prof. G. Caire Ph.D., from the Technical University of Berlin. His research interests include signal processing, wireless channel estimation, predictive quality of service and machine learning.
\end{IEEEbiography}
\begin{IEEEbiography} [{\includegraphics[width=1in,height=1.25in,clip,keepaspectratio]{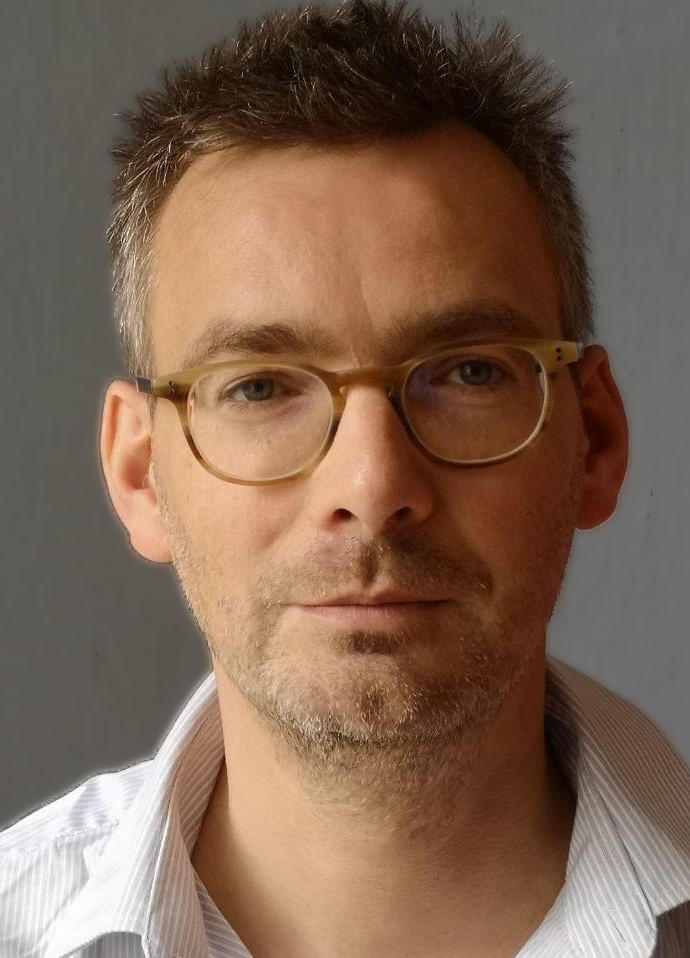}}]
    {Peter Jung} (Member IEEE, Member VDE/ITG) received
   the Dipl.-Phys. in high energy physics in 2000 from Humboldt
   University, Berlin, Germany, in cooperation with DESY Hamburg. Since
   2001 he has been with the Department of Broadband Mobile
   Communication Networks, Fraunhofer Institute for Telecommunications,
   Heinrich-Hertz-Institut (HHI) and since 2004 with Fraunhofer
   German-Sino Lab for Mobile Communications.  He received the
   Dr.-rer.nat (Ph.D.) degree in 2007 (on Weyl--Heisenberg
   representations in communication theory) at the Technical
   University of Berlin (TUB), Germany. P. Jung is working
   under DFG grants at TUB in the field of signal processing, information and communication
   theory and data science. In 2021 he was a visiting professor at TU Munich and associated with the Munich AI Future Lab (AI4EO).
   His current research interests are in the
   area compressed sensing, machine learning, time--frequency analysis, dimension reduction and randomized algorithms.  He is giving lectures in compressed sensing, estimation theory and inverse problems.
\end{IEEEbiography}
\begin{IEEEbiography}
    [{\includegraphics[width=1in,height=1.25in,clip,keepaspectratio]{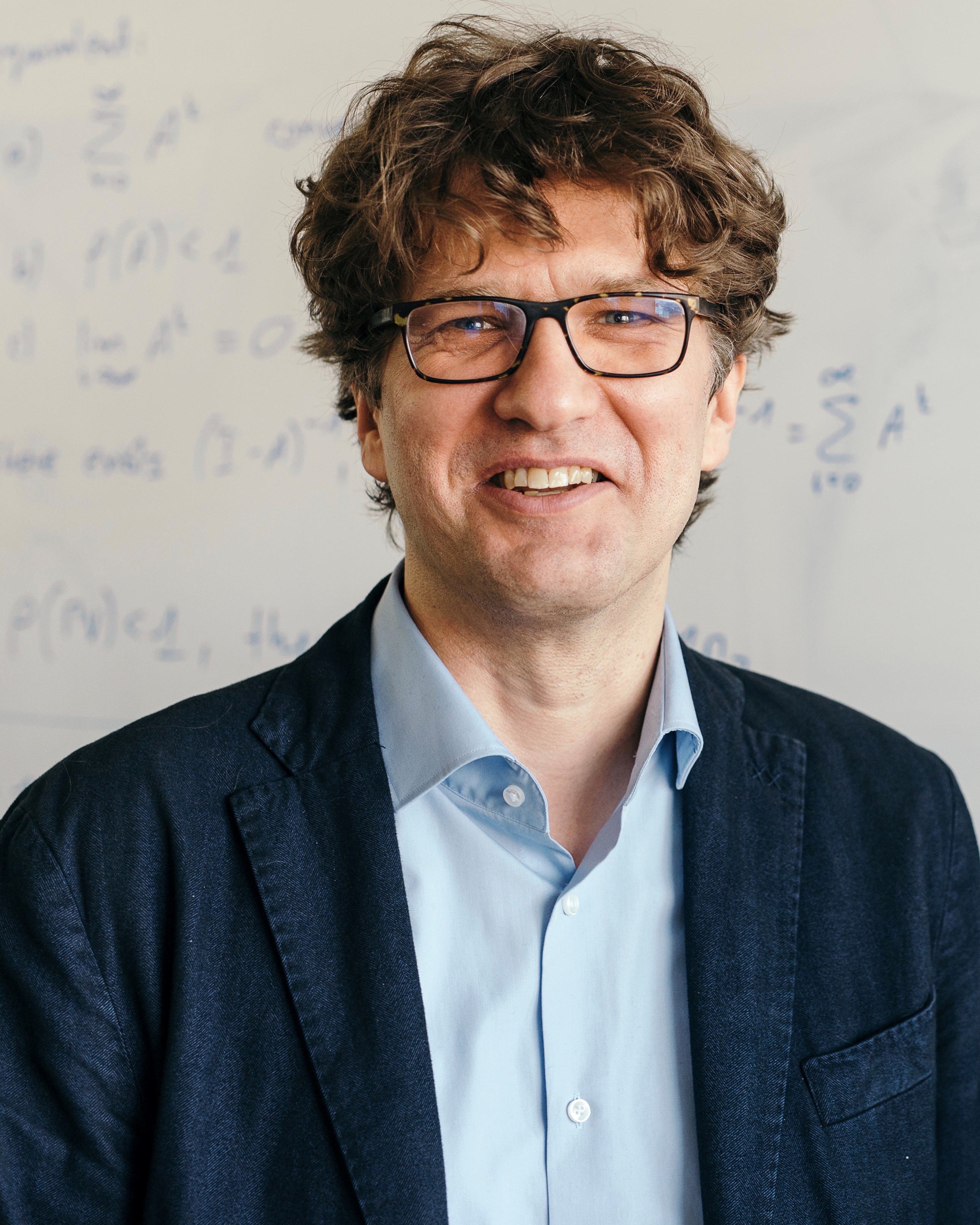}}]
    {SŁAWOMIR STAŃCZAK} 
    is Professor of Network Information Theory at the Technical University of Berlin and Head of the Wireless Communications and Networks Department at the Fraunhofer Heinrich Hertz Institute (HHI). Prof. Stanczak is co-author of two books and more than 200 peer-reviewed journal articles and conference papers in the field of information theory, wireless communications, signal processing, and machine learning. Prof. Stanczak received research grants from the German Research Foundation and the Best Paper Award from the German Society for Telecommunications in 2014. He was an associate editor of the IEEE Transactions on Signal Processing from 2012 to 2015 and chair of the ITU-T Focus Group on Machine Learning for Future Networks including 5G from 2017 to 2020. Since 2020 Prof. Stanczak is chairman of the 5G Berlin association and since 2021 he is coordinator of the projects 6G-RIC (Research \& Innovation Cluster) and CampusOS.
\end{IEEEbiography}
\end{document}

%% file: header.tex
\usepackage[dvipdfmx]{graphicx}
\usepackage{bmpsize}
\usepackage{url}
\usepackage{amsfonts}
\usepackage{balance}
\usepackage{multirow}
\usepackage{color}
\usepackage{cite}
\usepackage{epstopdf}
\makeatletter
\usepackage[caption=false]{subfig}
\let\MYcaption\@makecaption
\makeatother

\makeatletter
\let\@makecaption\MYcaption
\makeatother
\usepackage[T1]{fontenc}
\usepackage[scaled]{helvet}
\usepackage{algorithmicx}
\usepackage{algorithm}
\usepackage{amsmath}
\usepackage{amssymb}
\usepackage{algpseudocode}
\usepackage{bm}
\usepackage{enumitem}
\makeatletter
\let\MYcaption\@makecaption
\makeatother
\usepackage{xcolor,colortbl}
\usepackage{todonotes}
\usepackage{glossaries}
\usepackage{gensymb}
\usepackage{cleveref}
\usepackage{comment}
\usepackage{nicefrac}
\usepackage{textcomp}
\usepackage{pgfplotstable}
\usepackage{pgfplots}
\usepackage{xstring}
\usepackage{todonotes}
\usepackage{pifont}
\usepackage{etoolbox}
\usepgfplotslibrary{dateplot,statistics}
\usetikzlibrary{shapes,arrows.meta,calc}
\pgfplotsset{compat=1.15}
\usepackage{float}

\usepackage{mathtools}
\usepackage{bm}

\newcommand{\Z}{\ensuremath{\mathbb{Z}}}
\newcommand{\R}{\ensuremath{\mathbb{R}}}
\newcommand{\C}{\ensuremath{\mathbb{C}}}

\newcommand{\e}{\ensuremath{\mathrm{e}}}

\renewcommand*{\d}{\mathop{\kern0pt\mathrm{d}}\!{}}
\renewcommand{\vec}[1]{\ensuremath{\bm{#1}}}
\DeclareMathOperator{\vectorize}{vec}
\DeclareMathOperator*{\argmax}{argmax}

\crefname{equation}{}{}
\Crefname{equation}{Eq.}{Eqs.}
\crefname{figure}{Figure}{Figures}
\crefname{tabular}{Table}{Tables}
\crefname{table}{Table}{Tables}
\crefname{section}{Section}{Sections}

\algrenewcommand\algorithmicrequire{\textbf{Input}}
\algrenewcommand\algorithmicensure{\textbf{Output}}

\usepackage[per-mode=symbol,binary-units=true]{siunitx}
\DeclareSIUnit{\belmilliwatt}{Bm}
\DeclareSIUnit{\dBm}{\deci\belmilliwatt}

\input{gloss}
\usetikzlibrary{
    pgfplots.colorbrewer,
	patterns
}
\pgfkeys{
    /pgf/number format/.cd,
        set thousands separator={,},
        min exponent for 1000 sep=4,
}
\pgfplotsset{
    cycle list/.define={my marks}{
        every mark/.append style={solid,fill=\pgfkeysvalueof{/pgfplots/mark list fill}},mark=*\\
        every mark/.append style={solid,fill=\pgfkeysvalueof{/pgfplots/mark list fill}},mark=square*\\
        every mark/.append style={solid,fill=\pgfkeysvalueof{/pgfplots/mark list fill}},mark=triangle*\\
        every mark/.append style={solid,fill=\pgfkeysvalueof{/pgfplots/mark list fill}},mark=diamond*\\
    },
		cycle list/.define={my lines}{
			solid\\
			dotted\\
			densely dotted\\
			densely dashed\\
			dashdotted\\
			loosely dotted\\
			loosely dashed\\
			dashdotteddotted\\
    },
    cycle list/Accent-8,
	custom cycle list/.style={
				 legend style={font=\tiny},%
				 legend pos={south west},%
			   ,cycle list/Dark2-8,
			    mark list fill={.!75!white},
			    grid=both,
			    cycle multiindex* list={
			        Dark2-8
			            \nextlist
			        my marks
			            \nextlist
							my lines
			            \nextlist
			        very thick
			            \nextlist
			    }
	},
}
\usepgfplotslibrary{fillbetween}
\pgfplotsset{scaled x ticks=false}
\pgfplotsset{scaled y ticks=false}
\pgfplotsset{tick label style={/pgf/number format/fixed}}

\tikzset{
	>={Latex[width=2mm, length=2mm]},
	switch/.style= {rectangle, draw=blue!10!black, minimum width = 2cm, text centered, font=\sffamily,fill=blue!40!gray!30,ultra thick},
	base/.style = {switch, rounded corners},
	cond/.style = {switch, diamond,minimum width = 1cm,  aspect=2},
	connector/.style = {draw,ultra thick,->},
	ray/.style = {draw, -latex, dotted, thick, yellow, opacity=0.99},
	senses/.style = {draw, -latex, densely dotted, thick, magenta, opacity=0.5},
	communicates/.style = {draw, -latex, densely dotted, ultra thick,bend right=50},
}

%% file: gloss.tex
\newacronym{v2v}{V2V}{vehicle-to-vehicle}
\newacronym{v2x}{V2X}{vehicle-to-everything}
\newacronym{v2i}{V2I}{vehicle-to-infrastructure}
\newacronym{lte}{LTE}{long-term evolution}
\newacronym{los}{LOS}{line-of-sight}
\newacronym{nlos}{NLOS}{non line-of-sight}
\newacronym{ofdm}{OFDM}{orthogonal frequency-division multiplexing modulation}
\newacronym{otfs}{OTFS}{orthogonal time-frequency and space modulation}
\newacronym{otsm}{OTSM}{orthogonal time-sequency multiplexing}
\newacronym{sc-fdma}{SC-FDMA}{single-carrier frequency-division multiple access}
\newacronym{cir}{CIR}{channel impulse response}
\newacronym{fer}{FER}{frame error rate}
\newacronym{ici}{ICI}{inter-carrier interference}
\newacronym{ber}{BER}{bit error rate}
\newacronym{bler}{BLER}{block error rate}
\newacronym{siso}{SISO}{single input single output}
\newacronym{qos}{QoS}{quality of service}
\newacronym{CS}{CS}{compressed sensing}
\newacronym{ls}{LS}{least-squares}
\newacronym{fbmc}{FBMC}{filter bank multicarrier}
\newacronym{tf}{TF}{time-frequency}
\newacronym{dd}{DD}{delay-Doppler}
\newacronym{ds}{DS}{delay-sequency}
\newacronym{snr}{SNR}{signal-to-noise-ratio}
\newacronym{gfdm}{GFDM}{generalized frequency division multiplexing}
\newacronym{awgn}{AWGN}{additive white Gaussian noise} 
\newacronym{mmse}{MMSE}{minimum mean square error} 
\newacronym{lmmse}{LMMSE}{linear minimum mean square error} 
\newacronym{mle}{MLE}{maximum-likelihood estimator}
\newacronym{sir}{SIR}{signal-to-interference-ratio}
\newacronym{sirn}{SINR}{signal-to-interference-plus-noise-ratio}
\newacronym{cp}{CP}{cyclic prefix}
\newacronym{bfdm}{BFDM}{biorthogonal frequency division multiplexing}
\newacronym{wssus}{WSSUS}{wide-sense stationary uncorrelated scattering}
\newacronym{csi}{CSI}{channel state information}
\newacronym{ltv}{LTV}{linear time-varying}
\newacronym{cmd}{CMD}{channel main diagonal}
\newacronym{idg}{IDG}{isotropic discrete gradients}
\newacronym{fwht}{FWHT}{fast Walsh-Hadamard transform}
\newacronym{1D-fwht}{1D-FWHT}{1D fast Walsh-Hadamard transform}
\newacronym{2D-fwht}{2D-FWHT}{2D fast Walsh-Hadamard transform}
\newacronym{fft}{FFT}{fast Fourier transform}
\newacronym{dft}{DFT}{discrete Fourier transform}
\newacronym{dsft}{DSFT}{discrete symplectic Fourier transform}
\newacronym{1D-dsft}{1D-DSFT}{discrete symplectic Fourier transform}
\newacronym{2D-dsft}{2D-DSFT}{2D discrete symplectic Fourier transform}
\newacronym{qam}{QAM}{quadrature amplitude modulation}
\newacronym{qpsk}{QPSK}{quadrature phase shift keying}
\newacronym{5g}{5G}{5th generation wireless system}
\newacronym{4g}{4G}{4th generation wireless system}
\newacronym{isi}{ISI}{inter-symbol interference}
\newacronym{op}{OP}{orthogonal precoding}
\newacronym{od}{OD}{orthogonal decoding}
\newacronym{urllc}{URLLC}{ultra-reliable low-latency communication}
\newacronym{wifi}{WiFi}{wireless local area network}
\newacronym{opt}{OPT}{orthogonal precoding transform}
\newacronym{odt}{ODT}{orthogonal decoding transform}
\newacronym{cmde}{CMDE}{channel main diagonal estimation}
\newacronym{sf}{SF}{sub-frames}
\newacronym{srh}{SRH}{smoothness regularized Hessian}
\newacronym{srh-mna}{SRH-MNA}{smoothness regularized Hessian with mode- and noise-awareness}
\newacronym{srh-ma}{SRH-MA}{smoothness regularized Hessian with mode-awareness}
\newacronym{srh-na}{SRH-NA}{smoothness regularized Hessian with noise-awareness}
\newacronym{mse}{MSE}{mean square error}
\newacronym{se-papr}{NMSED}{normalized maximal symbol-error deviation}

%% file: authors_conf.tex
\title{\LARGE Estimation of Doubly-Dispersive Channels in Linearly Precoded Multicarrier Systems Using Smoothness Regularization}   %
\author{

\IEEEauthorblockN{
				Andreas Pfadler\IEEEauthorrefmark{1}\IEEEauthorrefmark{3},
				Tom Szollmann\IEEEauthorrefmark{1}\IEEEauthorrefmark{3},
				Peter Jung\IEEEauthorrefmark{2}\IEEEauthorrefmark{3}
				and Slawomir Stanczak\IEEEauthorrefmark{2}\IEEEauthorrefmark{3}
}
\IEEEauthorblockA{\IEEEauthorrefmark{1}Volkswagen Commercial Vehicles, Wolfsburg, Germany\\ \{andreas.pfadler, tom.szollmann\}@volkswagen.de}
\IEEEauthorblockA{\IEEEauthorrefmark{2}Fraunhofer Heinrich Hertz Institute, Berlin, Germany\\ \{peter.jung, slawomir.stanczak\}@hhi.fraunhofer.de}
\IEEEauthorblockA{\IEEEauthorrefmark{3}Technical University of Berlin, Berlin, Germany}
}